\documentclass[pra,twocolumn,superscriptaddress,amsmath,amssymb,floatfix,noshowpacs]{revtex4}
\usepackage{graphicx,hyperref}
\usepackage{color}
\usepackage{soul}
\usepackage{amsmath}
\usepackage{tabularx}
\usepackage{array}
\usepackage{xcolor}
\usepackage[toc,page]{appendix}
\usepackage{multirow}

\graphicspath{{fig/}}

\newcommand*{\ket}[1]{\vert #1 \rangle}

\newcommand{\hide}[1]{}

\begin{document}
\title{\textbf{Universal approach for quantum interfaces with atomic arrays}}
\author{Yakov Solomons}
\affiliation{Department of Chemical \& Biological Physics, Weizmann Institute of Science, Rehovot 7610001, Israel}
\author{Roni Ben-Maimon}
\affiliation{Department of Chemical \& Biological Physics, Weizmann Institute of Science, Rehovot 7610001, Israel}
\author{Ephraim Shahmoon}
\affiliation{Department of Chemical \& Biological Physics, Weizmann Institute of Science, Rehovot 7610001, Israel}
\date{\today}

\begin{abstract}
We develop a general approach for the characterization of atom-array platforms as light-matter interfaces, focusing on their application in quantum memory and photonic entanglement generation. Our approach is based on the mapping of atom-array problems to a generic 1D model of light interacting with a collective dipole. We find that the efficiency of light-matter coupling, which in turn determines those of quantum memory and entanglement, is given by the on-resonance reflectivity of the 1D scattering problem, $r_0=C/(1+C)$, where $C$ is a cooperativity parameter of the model. For 2D and 3D atomic arrays in free space, we derive the mapping parameter $C$ and hence $r_0$, while accounting for realistic effects such as the finite sizes of the array and illuminating beam and weak disorder in atomic positions.
Our analytical results are verified numerically and reveal a key idea: efficiencies of quantum tasks are reduced by our approach to the classical calculation of a reflectivity. This provides a unified framework for the analysis of collective light-matter coupling in various relevant platforms such as optical lattices and tweezer arrays. Generalization to collective systems beyond arrays is discussed.
\end{abstract}
 \maketitle
\section{Introduction}
Quantum optical platforms, based on the manipulation of atoms and photons, play an essential role in the exploration of quantum science and technology. Of crucial importance is the ability to establish an interface between photons and atoms. Such an interface allows to benefit from the low-loss propagation of photons combined with the quantum coherence or nonlinearity of atoms, with applications ranging from quantum memories and information to many-body physics \cite{Lvovsky,coop_Chang}. To this end, a quantum interface is required to couple a certain target photon mode which one excites and detects, to a relevant spatially matched atomic degree of freedom. In turn, the efficiency of the interface is characterized by the ratio between the emission rate of the atomic degree of freedom to the target mode and that to the rest of the undesired modes. This ratio depends on the specific realization: For an ensemble of atoms trapped in free space \cite{Liu,Phillips,Julsgaard,Eisaman,Chaneliere,FIM,Choi,GorshkovFree1,GorshkovFree2,POLr,Ofer} or along a waveguide \cite{Gouraud,Sayrin} it is typically given by the so-called optical depth, whereas for atoms trapped inside a cavity this ratio is often identical to the cooperativity parameter \cite{Cirac,GorshkovCavity,Specht,Ritter,Giannelli,Korber}.

Recently, spatially ordered arrays of trapped atoms, as can be realized in an optical lattice \cite{Bloch1,Bloch2}, have emerged as a novel quantum light-matter interface \cite{Facchinetti,Bettles_mirror,Efi5,Manzoni,Grankin,Guimond,Henriet,Efi1,Efi2,Efi3,Parmee,Asenjo-Garcia,Masson,Patti,Castells,Perczel,Bettles,Plankensteiner,Efi4,Rui,MALZ,ALJ1,POHs}. For arrays with near-wavelength lattice spacing, the combination of spatial order with the collective response of the atoms to light
considerably reduces the scattering into unwanted directions. This results in strong and directional light-matter coupling between a propagating, target photon mode and a spatially matched collective dipole of the atoms. In 2D arrays, the strength of the coupling is evident and characterized by the high reflectivity of the target mode scattered off the array \cite{Bettles_mirror,Efi5,Rui}. Moreover, the reflectivity seems to appear in relation to efficiencies of various quantum applications that were subsequently proposed, from quantum entanglement generation and information with photons \cite{Bekenstein,Wei,Moreno,Zhang,Our_paper,Srakaew}, to optomechanics \cite{Efi2,Efi3} and quantum memories \cite{Facchinetti,Manzoni}. However, no clear relation or framework that underscores the general role of array reflectivity in such light-matter applications was established thus far.

In this work, we provide a general approach for the analysis and characterization of atom-array quantum interfaces. Our strategy is based on the mapping of various atom-array light-matter problems to a simple 1D model of photons scattering off a dipole. The model is fully characterized by the resonant reflectivity $r_0$, which is found to be related to the cooperativity parameter $C$ via $r_0=C/(1+C)$. Considering quantum light-matter applications, such as quantum memory or photonic entanglement generation, we find their efficiencies to be given by $r_0$. This mirrors the energy balance of the underlying classical 1D scattering problem, hence the reflectivity $r_0$ is expected to characterize other quantum tasks as well. Then, for the consideration of such quantum applications in atom-array platforms, one merely has to map these systems to the 1D model, where calculations are simple.

We illustrate this idea by considering relevant examples of 2D and 3D atomic arrays as can be realized in optical lattices \cite{Bloch1,Bloch2} or tweezer-arrays \cite{tweezer1,tweezer2}. We establish their mapping to the 1D model, finding $C$ and the reflectivity $r_0$. Special care is taken to the collective nature of their interaction with light (dipole-dipole interactions) which is key for efficient light-matter coupling in arrays, while also accounting for realistic finite-size effects and imperfections. The reflectivity $r_0$, extracted analytically from the mapping, is verified by its agreement with the full numerical calculation of the scattering.
The appealing idea, that a simple classical calculation of reflectivity is sufficient to determine efficiencies of quantum tasks, could be extended to more complex and general scenarios provided that their mapping to the generic 1D model is established. To this end, we discuss general considerations for this mapping in collective many-atom systems.

\section{Summary of results}
We begin with a brief account of the results and ideas presented in this work. In particular, Table I below summarizes results derived for specific atom-array interfaces. Prospects and extensions are discussed in Sec. VIII.
\subsection{Reflectivity as a figure of merit of a quantum interface: 1D model (Sec. III)}
We introduce a generic 1D scattering model which serves as a minimal model for a quantum atom-photon interface, see Fig. 1 and Eq. (\ref{model}):  A collective atomic dipole $\hat{P}$ is predominantly coupled to a 1D-propagating ``target" photon mode $\hat{\mathcal{E}}$, while also scattering to other modes considered as a loss. The ratio of these couplings, described by the cooperativity $C=\Gamma/\gamma_{\mathrm{loss}}$, characterizes the model and is found to be related to the on-resonance reflectivity of the target mode, $r_0=C/(C+1)$. We show that the reflectivity $r_0$ is equal to the efficiency of power conversion between the dipole $\hat{P}$ and the target mode $\hat{\mathcal{E}}$ and hence emerges as the efficiency of quantum tasks. In particular, by implementing schemes of a quantum memory and photon-photon entanglement within the 1D model we show that $r_0$ is the efficiency of both. Therefore, for any quantum-interface system that can be mapped to the 1D model, the reflectivity $r_0$ serves as the universal figure of merit. The latter is obtained from $C$ or by a simple classical calculation of the linear reflection of the target mode off the atoms.
\subsection{Mapping to the 1D model: general considerations (Sec. IV)}
For a given system wherein collective light-matter effects are accounted for, the first challenge is to identify the mapping to the 1D model. Considering a general many-atom system in free-space and including the photon-mediated dipole-dipole interactions between the atoms, we find a simple sufficient condition for the mapping. Namely, the target mode $\hat{\mathcal{E}}$ should spatially match with a single eigenmode, or a single approximate eigenmode, $\hat{P}$ of the dipole-dipole interaction kernel.
\begin{figure}[t!]
  \begin{center}
    \includegraphics[width=\columnwidth]{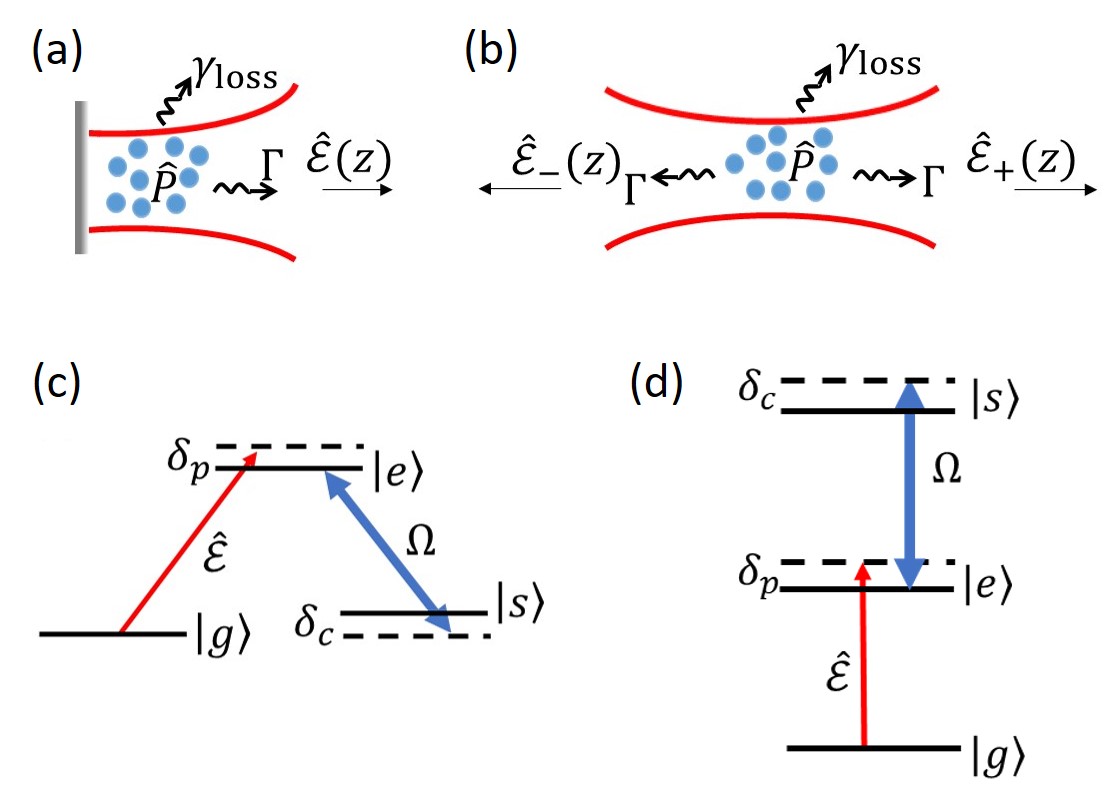}
    \caption{\small{Generic 1D model of a light-matter interface [Eqs. \eqref{model}] onto which various systems are mapped. (a)  A collective dipole $\hat{P}$ is coupled to the target mode $\hat{\mathcal{E}}$ at an emission rate $\Gamma$, while scattering to other modes (loss) at rate $\gamma_{\mathrm{loss}}$. (b) Symmetric, two-sided version of the model [Eq. \eqref{inout}]. (c) Level structure of the atoms in the three-level variant of the model used to describe a quantum memory [Eqs. \eqref{model2}]. (d) Ladder-configuration version of three-level atoms.
    }} \label{fig1}
  \end{center}
\end{figure}
\subsection{Atom-array interfaces (Secs. V, VI, VII)}
The mapping to the 1D model becomes rather natural upon the consideration of atoms placed in an ordered array. Then, translational lattice symmetry dictates that the dipole eigenmodes are lattice Fourier modes. This holds to a very good approximation also for a finite-size array if the number of atoms is large enough and the target mode is paraxial. In this case the relevant dipole eigenmodes still resemble Fourier modes, and the paraxial target mode, e.g. a Gaussian beam, spatially matches a Gaussian-shaped dipole mode which constitutes an approximate dipole eigenmode (being a superposition of quasi-degenerate, low-spatial-frequency dipole eigenmodes).

We demonstrate this approach by establishing the mapping of several relevant cases of atomic arrays, as summarized in Table I. We begin in Sec. V by considering a finite-size 2D array illuminated by paraxial light (e.g. a Gaussian beam), also accounting for weak disorder and the possibility of lattice spacing exceeding the wavelength (e.g. as typical of tweezer arrays). The reflectivity $r_0$, found analytically from the mapping (Tables I and II), is compared to that obtained by a full numerical calculation of the classical scattering off the finite-size and weakly disordered array, finding excellent agreement (Fig. 4). We then extend the analysis for a 3D array comprised of a finite number of $N_z$ layers of 2D arrays (Sec. VI). We find that the mapping of a phased-matched array, whose inter-layer separation equals an integer (or half-integer) multiple of a wavelength, can exhibit a cooperativity enhanced by a factor $N_z$ (Table I). Finally, in Sec. VII we consider a spatially modulated 2D array (optical superlattice) wherein a quantum memory is realized by relying on collective dark states and without the need of three-level atoms \cite{Yelin}. Including finite-size effects and imperfections we again establish the mapping to the 1D model finding the memory efficiency $r_0$ identical to that of the usual, unmodulated 2D array from Sec. V and Table I. This highlights the universality of the reflectivity $r_0$ as a figure of merit, here of quantum memory, independently of the specific protocol or scheme considered.

\begin{table}[t!]
\centering
\begin{tabular}{| >{\centering\arraybackslash}m{7em} | >{\centering\arraybackslash}m{9em} | >{\centering\arraybackslash}m{9em} |}
 \hline
 & Emission rate to target mode $\quad \quad \quad\Gamma$& Emission rate to undesired modes $\gamma_{\mathrm{loss}}$\\ [0.5ex]
 \hline\hline
2D atom array $a<\lambda$ (Sec. V) & \multirow{2}{*}{$\eta\Gamma_0=\eta\frac{3}{4\pi}\frac{\lambda^2}{a^2}\gamma$} &  $(1-\eta)\Gamma_0 + \gamma_s$ \\
\cline{1-1} \cline{3-3}
2D atom array $a>\lambda$ (Sec. V) & &  $(1-\eta)\Gamma_0 + \gamma_s+\gamma_{\mathrm{diff}}$ \\
\hline
3D atom array $a<\lambda$ (Sec. VI) & $\eta N_z\Gamma_0=\eta\frac{3}{4\pi}\frac{\lambda^2}{a^2}N_z\gamma$ &  $(1-\eta)N_z\Gamma_0 + \gamma_s$ \\
\hline
Cavity (reference case) & $\frac{3}{4\pi}\frac{\lambda^2}{\pi w^2/2}\frac{\mathcal{F}}{\pi}N\gamma$ & $\gamma$\\
\hline
\end{tabular}
\caption{Mapping of atom-array interfaces to the generic 1D model of Fig. 1 and Sec. III (model parameters $\Gamma$ and $\gamma_{\mathrm{loss}}$).
For a 2D array ($N$ atoms, lattice constant $a$) the collective coupling rate $\Gamma_0$ to the target mode is reduced by the overlap factor $\eta=\mathrm{erf}^2(\frac{a\sqrt{N}}{\sqrt{2}w})<1$ between the target-mode profile (waist $w$) and the finite size array. Here $\gamma$ is the individual-atom free-space spontaneous emission rate. For lattice spacing exceeding the wavelength, $a>\lambda$, losses due to scattering to higher diffraction orders occur [$\gamma_{\mathrm{diff}}$, Eq. (\ref{diff})], in addition to the individual-atom losses ($\gamma_s$, e.g. due to disorder) and those due to non-perfect spatial overlap [$(1-\eta)\Gamma_0$]. Moving to 3D arrays, phase-matching conditions enhance collective emission by a factor $N_z$ (the number of layers, Fig. \ref{3D_array}): for a good overlap $\eta\rightarrow1$, the cooperativity $C=\Gamma/\gamma_{\mathrm{loss}}$ also increases by $N_z$. Comparing these free-space array cases to a dilute atomic ensemble inside a cavity (e.g. Ref. \cite{GorshkovCavity}), we observe a similar scaling of the coupling $\Gamma$ (noting that $N/w^2$ is the effective 2D density in analogy to $N_z/a^2$ of the array), but with an enhancement of $\mathcal{F}$ from the cavity finesse. The cavity enhancement $\mathcal{F}$ is required for a dilute ensemble to combat its large losses arising from individual-atom transverse scattering out of the cavity at a rate $\gamma\sim\Gamma_0$ [compared to much smaller losses of ordered arrays, $(1-\eta)\Gamma_0$, $\gamma_s$, see Table \ref{table:2}].}
\label{table:1}
\end{table}
\section{Generic 1D model: cooperativity and reflectivity}
We begin by introducing a generic 1D-like scattering problem which captures the essence of an atom-photon interface.
Using a classical optics perspective, we first highlight the meaning of and the relation between cooperativity, efficiency and reflectivity. We then relate them to the problems of quantum memory and photon entanglement generation, showing that the efficiencies of the latter are identical to the reflectivity. This leads to our universal approach for quantum interfaces, consisting of the mapping of collective light-matter problems to the 1D scattering model.

\subsection{The model}
We consider a paraxial mode of light with some finite mode area, e.g. a Gaussian beam within its Rayleigh range, propagating along $z$ and described by the 1D field operator $\hat{\mathcal{E}}(z,t)$, with $[\hat{\mathcal{E}}(z),\hat{\mathcal{E}}^{\dag}(z')]\propto\delta(z-z')$. This ``target mode" interacts with a collection of two-level atoms and couples to a spatially matched collective dipole of the atoms, described by the operator $\hat{P}$ (Fig. \ref{fig1} a,b). However, since this dipole is comprised of discrete atoms, it may in general also scatter to other undesired photonic modes in different directions. A simple model that captures this scenario is given by
\begin{eqnarray}
\frac{d\hat{P}}{dt}&=&\left[i(\delta_p-\Delta)-\frac{\Gamma+\gamma_{\mathrm{loss}}}{2}\right]\hat{P}+i\sqrt{\Gamma}\hat{\mathcal{E}}_0(0,t)+\hat{F}(t),
\nonumber\\
\hat{\mathcal{E}}(z)&=&\hat{\mathcal{E}}_0(z)+i\sqrt{\Gamma}\hat{P}.
\label{model}
\end{eqnarray}
These Heisenberg-picture equations describe a 1D scattering problem of the target mode $\hat{\mathcal{E}}(z)$ off the dipole $\hat{P}$. Here, $\Gamma$ is the emission rate of the atomic dipole to the target-mode 1D continuum, with an input quantum field satisfying $[\hat{\mathcal{E}}_0(0,t),\hat{\mathcal{E}}_0^{\dag}(0,t')]=\delta(t-t')$, whereas $\gamma_{\mathrm{loss}}$ is the emission rate to the undesired modes with corresponding quantum vacuum noise $\langle\hat{F}(t)\hat{F}^{\dag}(t')\rangle=\gamma_{\mathrm{loss}}\delta(t-t')$. $\delta_p=\omega_p-\omega_a$ is the detuning between the central frequency of the target mode ($\omega_p$) and that of the ``bare" two-level atom transition ($\omega_a$), whereas $\Delta$ is a possible collective shift of the collective atomic dipole.

\subsection{Cooperativity and coupling efficiency}
We define the cooperativity $C$ as the branching ratio between the emission to the desired target mode and that to undesired modes,
\begin{eqnarray}
C=\frac{\Gamma}{\gamma_{\mathrm{loss}}}.
\label{C}
\end{eqnarray}
Quantum mechanically, this is a ratio of spontaneous emission rates, but it also has a clear classical meaning as the ratio of radiated energy or power. In Appendix A we illustrate this point in two ways. First, we consider an initially excited dipole and calculate classically the fraction of energy radiated into the target mode, finding
\begin{eqnarray}
r_0\equiv\frac{C}{C+1}.
\label{r_0}
\end{eqnarray}
Second, we consider an input continuous wave (CW) illumination in the target mode and calculate the fraction of the power absorbed by the dipole $\hat{P}$ in steady state. At resonance $\delta_p=\Delta$ we again obtain $r_0$ from Eq. (\ref{r_0}). Therefore, $r_0$ describes the light-matter coupling efficiency between the target mode and the atoms. It originates in the underlying classical linear optics problem and is determined by $C$ (as also noted before in the context of various quantum optical platforms \cite{coop_Chang,coop_Chang2,coop_Lodahl}).

\subsection{Reflectivity as an efficiency}
The coupling efficiency from Eq. \eqref{r_0} can be interpreted as the reflectivity of the 1D scattering problem. To show this we first note that the model from Eqs. (\ref{model}) is valid for either a one-sided scattering problem (Fig. \ref{fig1}a) or for a two-sided problem (Fig. \ref{fig1}b). We now consider the latter, for which the field is separated into right- and left-propagating components, $\hat{\mathcal{E}}_{\pm}(z)$. The input field in the equation for $\hat{P}$ is then understood as the symmetric superposition of both sides, $\hat{\mathcal{E}}_0(z)=\frac{1}{\sqrt{2}}[\hat{\mathcal{E}}_{0,+}(z) +\hat{\mathcal{E}}_{0,-}(-z)]e^{-i k_pz}$ ($k_p\equiv\omega_p/c$, $z>0$), and the equation for the output field of each component is
\begin{eqnarray}
\hat{\mathcal{E}}_{\pm}(z)&=&\hat{\mathcal{E}}_{0,\pm}(z)+i\sqrt{\frac{\Gamma}{2}}e^{\pm i k_p z}\hat{P}.
\label{inout}
\end{eqnarray}
For classical CW light shined in either direction, we find the amplitude reflectivity as (Appendix A)
\begin{eqnarray}
r=-\frac{\Gamma}{\Gamma+\gamma_{\mathrm{loss}}+i2(\Delta-\delta_p)} \quad\rightarrow \quad r_0=|r(\delta_p=\Delta)|.
\label{r}
\end{eqnarray}
We thus identify that the magnitude of this \emph{field} reflectivity at resonance, $r_0$, is equal to the efficiency associated with \emph{power} conversion from Eq. \eqref{r_0}. For the intensity (power) reflectivity, we thus have $R=r_0^2$ which lands itself to an intuitive two-step picture of reflectivity: First, the photons of the target mode are absorbed by the atoms at efficiency $r_0$, as discussed in the previous subsection, and then they are emitted from the atoms to the target mode also at efficiency $r_0$, leading to $R=r_0^2$. While in a CW scenario this two-step process never really occurs at a certain timing, this simple picture is useful to understand the physics of the quantum memory.

\subsection{Quantum memory efficiency}
Consider now that the atomic system is comprised of three-level atoms, with a stable level $|s\rangle$ in addition to the ground state $|g\rangle$ (Fig. \ref{fig1}c). Then, there exists another relevant collective atomic spin, $\hat{S}$, which accounts for the coherence between these two stable levels, and that is coupled to $\hat{P}$ via an external field $\Omega$. Eqs. (\ref{model}) are now modified to
\begin{eqnarray}
\frac{d\hat{P}}{dt}&=&\left[i(\delta_p-\Delta)-\frac{\Gamma+\gamma_{\mathrm{loss}}}{2}\right]\hat{P}+i\Omega\hat{S}+i\sqrt{\Gamma}\hat{\mathcal{E}}_0(0,t)\nonumber\\&+&\hat{F}(t),
\nonumber\\
\frac{d\hat{S}}{dt}&=&i\delta_2\hat{S}+i\Omega^{\ast}\hat{P},
\nonumber\\
\hat{\mathcal{E}}(z)&=&\hat{\mathcal{E}}_0(z)+i\sqrt{\Gamma}\hat{P},
\label{model2}
\end{eqnarray}
where $\delta_2=\delta_p-\delta_c$ is the combined detuning of the two-photon transition from $|g\rangle$ to $|s\rangle$ (Fig. 1c), and the emission rate to undesired modes $\gamma_{\mathrm{loss}}$ and the corresponding quantum vacuum noise $\hat{F}(t)$, may now include contributions from processes involving the third level.

The goal of a quantum memory is to coherently transfer the excitations and the quantum state of a pulse of the target mode $\hat{\mathcal{E}}$ into the stable spin $\hat{S}$. A second goal is to be able to retrieve these excitations and the quantum state from $\hat{S}$ back to the propagating mode of $\hat{\mathcal{E}}$. Within this process, the dipole $\hat{P}$ mediates the interaction between the field and the stable spin $\hat{S}$, via the tunable coupling $\Omega(t)$.

We analyzed the quantum memory protocol for the model in Eqs. \eqref{model2}. The analysis is performed in close analogy to that of Ref. \cite{GorshkovCavity} for an atomic ensemble inside an optical cavity, by noting that the latter is equivalent to Eqs. \eqref{model2} in the fast cavity regime where the cavity mode couples the atoms to a spatially matched output mode. We present the details of this analysis in Appendix B, including the expressions of the control-pulse temporal shape $\Omega(t)$ needed for optimal storage (retrieval), given the temporal shape of the input (output) photon wave packet one wishes to store (retrieve). We obtain that the optimal storage and retrieval efficiencies are both equal to $r_0=C/(C+1)$, reflecting that these are time-reversed processes \cite{GorshkovCavity}. The meaning, revealed by our approach, is that the on-resonance reflectivity $r_0=|r(\delta_p=\Delta)|$ is in fact equal to the quantum memory efficiency. As shown in the subsection above, this in turn stems from the light-matter coupling efficiency of the underlying scattering problem. In particular, the  total efficiency of the quantum memory is given by the multiplication of the storage and retrieval efficiencies, $r_0^2$, which is nothing but the intensity reflectivity. Thus, the memory protocol, using the temporal tunability of the coupling field $\Omega(t)$, allows to break the reflection process into two distinct stages: absorption (storage) and re-emission (retrieval). But the combined effect is equivalent to intensity reflection, as exhibited by the total coupling efficiency $r_0^2$.
\subsection{Quantum correlation efficiency}
We now consider an extension of the model where nonlinearity of the atomic system is introduced and leads to effective photon-photon interactions and correlations useful e.g. in photonic quantum gates \cite{coop_Chang}. In its simplest form, such an interaction mechanism can be effectively modeled by allowing only a single excitation in the atomic system, truncating all multiply-excited states. For a single atom, this amounts to a two-level atom model, whereas for an atomic ensembles, e.g. an array, a similar situation is achieved via Rydberg blockade \cite{Lukin,Ofer_Ryd}. In the latter, the third atomic level of each atom, $\ket{s}$, is taken as a highly excited Rydberg state, forming a ladder system with $\delta_2=\delta_p+\delta_c$ (Fig. \ref{fig1}d). For strong enough interaction between Rydberg states of different atoms (large blockade radius \cite{Ofer_Ryd,Petrosyan}), all multiply-excited states of the form, $(\hat{S}^\dagger)^n\ket{gg...g}$ ($n>1$), are far-detuned and truncated \cite{Our_paper}.

Adiabatically eliminating $\hat{P}$ from Eq. \eqref{model2}, we obtain the dynamical equation for $\hat{S}$, as:
\begin{equation}
\frac{\partial \hat{S}}{\partial t}=-\bigg[\frac{\Gamma_S}{2}+i(\Delta_S-\delta_{2})\bigg]\hat{S}-\hat{\Omega}_S.
\end{equation}
where $\Gamma_S$, $\Delta_S$, and $\hat{\Omega}_S$ are the effective width, shift, and input field, of the collective two-photon transition from $\ket{g}$ to $\ket{s}$, given in Appendix B. Together with the blockade condition, $(\hat{S}^\dagger)^n\ket{gg...g}=0$, ($n>1$), this amounts to a scattering problem off a singe two-level emitter in 1D. Following Ref. \cite{Our_paper}, where the specific problem of a formally  infinite 2D array was treated,  we solve this general nonlinear problem analytically: Given a coherent-state input, we find the steady-state photon correlations of the output light. For example, at the resonances $\delta_2=0$ and $\delta_p=\Delta$, we obtain antibunching for the transmitted light, with $g^{(2)}(0)=|1-r_0^2|^2$. This expression clearly demonstrates that it is again the reflectivity of the array that determines the amount of correlations, as was also suggested by the numerical results of Refs. \cite{Moreno,Zhang} where the utility of such correlations for two-qubit gates was discussed. In Ref. \cite{Our_paper} a similar conclusion was reached also for the entanglement generated between different transverse modes scattered off a 2D array. Here we see however that these results originate in a generic feature of the nonlinear 1D model and should hence apply to any problem that can be mapped to it; Namely, the efficiency of the classical linear problem determines the entanglement generation of the quantum nonlinear problem it underlies, and these are given by the reflectivity $r_0$. We expect this general idea to extend to various mechanisms of array nonlinearity: e.g. in Ref. \cite{Efi2}, where array nonlinearity originates only from an optomechanical effect, the generated photon correlations were also shown to scale as the reflectivity.
\subsection{General approach for a quantum interface}
The considerations presented above establish that the on-resonance reflectivity $r_0$, along with the cooperativity $C$, form the figure of merit of the generic 1D light-matter interface. This is demonstrated for applications in quantum memories and quantum nonlinear optics and may be extended to other quantum tasks. This idea has an important practical meaning: For any given two-sided system that is mapped onto the generic 1D model of Eqs. \eqref{model} [or \eqref{model2}], one merely has to calculate the reflectivity of the underlying classical optics problem, in order to obtain efficiencies of relevant quantum problems.

In the following, we first discuss some general considerations of the mapping of collective atom-photon interfaces to the generic 1D model (Sec. IV) and then  demonstrate it for realistic cases of 2D and 3D atom-array interfaces (Secs. V-VII). In each case we identify the relevant collective atomic dipole $\hat{P}$ and photonic target mode $\hat{\mathcal{E}}$, perform the mapping onto the generic 1D model, and find analytically the effective model parameters $\Gamma$ and $\gamma_{\mathrm{loss}}$, from which the cooperativity $C$ is determined. These analytical results are summarized in Tables I and II. Comparison to numerical, classical-scattering calculations of the reflectivity shows excellent agreement (Fig.\ref{Sub_numerics}), thus illustrating the power and validity of our approach.

\section{Mapping a collective system to the 1D model}
We concluded above that the operation of any atom-photon system as a quantum interface is universally characterized by its reflectivity, provided that the system can be mapped to the generic 1D model. To establish this mapping, one first has to find the relevant collective dipole $\hat{P}$ and target mode $\hat{\mathcal{E}}$. In particular, this can become challenging for systems where the collective response of atoms to light is considered (dipole-dipole interactions). Here we begin with a full general model of a collection of atoms interacting with free-space photons modes. We derive Heisenberg-Langevin equations of motion and identify the coupling between collective dipole eigenmodes, and a paraxial target photon mode. It is seen that the mapping to the 1D model is possible if the target mode spatially matches a single dipole eigenmode or a superposition of quasi-degenerate dipole eigenmodes.

\subsection{System}
We consider a collection of identical three-level atoms (Fig. \ref{fig1}c), $n=1,...,N$, situated at fixed positions $\mathbf{r}_{n}=(\mathbf{r}^{\perp}_{n},z_n)$ ($\mathbf{r}_{\perp}$ and $z$ denoting projections along transverse $xy$ and longitudinal $z$ directions). The atoms are illuminated by a quantum field with central frequency $\omega_p$ working on the $\ket{g}\leftrightarrow\ket{e}$ transition, while an external coherent field $\Omega$ with frequency $\omega_c$ couples the levels $\ket{s}\leftrightarrow\ket{e}$ (Fig. 1c). The full Hamiltonian is given by
\begin{equation}\label{Seq1}
    \hat{H}=\hat{H}_f+\hat{H}_s+\hat{H}_I.
\end{equation}
Here  $\hat{H}_f=\sum_{\mathbf{k}_{\perp}}\sum_{k_z}\sum_{\mu}\hbar\omega_{\mathbf{k}_{\perp}k_z}\hat{a}^{\dagger}_{\textbf{k}_\perp k_z\mu}\hat{a}_{\textbf{k}_\perp k_z\mu}$ is the Hamiltonian of the photons in free space, characterized by the transverse and longitudinal wave-vectors $\mathbf{k}_{\perp}=(k_x,k_y)$ and $k_z$, respectively, the polarization index $\mu$, and $\omega_{\mathbf{k}_{\perp}k_z}=c\sqrt{|\mathbf{k}_{\perp}|^2+k_z^2}$. The Hamiltonian of the atomic system is given by
\begin{equation}
\hat{H}_s=\sum_n\bigg[\hbar\omega_{e}\hat{\sigma}_{ee,n}+\hbar\omega_{s}\hat{\sigma}_{ss,n}\bigg]-\bigg[\hbar\Omega e^{-i\omega_ct}\hat{\sigma}_{se,n}^\dagger+\mathrm{h.c}\bigg],
\label{H_s}
\end{equation}
where $\hat{\sigma}_{\alpha\alpha',n}=|\alpha\rangle_n\langle \alpha'|$ is the transition operator of  an atom $n$ and $\hbar\omega_{\alpha}$ is the energy of level $|\alpha\rangle$. The interaction Hamiltonian between the quantum field and the atoms in the dipole approximation reads
\begin{equation}
\hat{H}_I=-\sum_n\left[\mathbf{d}_{ge}\hat{\sigma}_{ge,n}+\mathbf{d}_{se}\hat{\sigma}_{se,n}+\mathrm{h.c.}\right]\cdot
\left[ \hat{\mathbf{E}}(\mathbf{r}^{\perp}_n,z_n)+ \mathrm{h.c.}\right],
\label{H_I}
\end{equation}
where $\mathbf{d}_{ge}=d\mathbf{e}_{d}$ is the dipole matrix element corresponding to the $\ket{g}\leftrightarrow \ket{e}$ transition and likewise $\mathbf{d}_{se}$ corresponds to the $\ket{s}\leftrightarrow \ket{e}$ transition. The quantum field operator is given by
\begin{equation}
\begin{split}
&\hat{\mathbf{E}}(\mathbf{r}_{\perp},z,t)=\\&=i\sum_{\mathbf{k}_{\perp}}\sum_{k_z}\sum_{\mu}\sqrt{\frac{\hbar\omega_{\mathbf{k}_{\perp}k_z}}{2\epsilon_0L^3}}\mathbf{e}_{\mathbf{k}_{\perp}k_z\mu}\hat{a}_{\mathbf{k}_{\perp}k_z\mu}(t)e^{i(\mathbf{k}_{\perp}\cdot\mathbf{r}_{\perp}+k_zz)},
\label{SE}
\end{split}
\end{equation}
with $L^3$ being the quantization volume and $\mathbf{e}_{\mathbf{k}_{\perp}k_z\mu}$ the photon polarization vector.

\subsection{Heisenberg-Langevin formalism}
Writing the Heisenberg equation for the photons  in the laser-rotated frame ($\hat{\sigma}_{ge,n}e^{i\omega_pt}\rightarrow\hat{\sigma}_{ge,n}$, $\hat{\sigma}_{gs,n}e^{i(\omega_p-\omega_c)t}\rightarrow\hat{\sigma}_{gs,n}$), we solve for the photon field under the Born-Markov approximation, obtaining
\begin{equation}
\hat{E}(\mathbf{r}_{\perp},z)=\hat{E}_{0}(\mathbf{r}_{\perp},z)+\frac{\omega_{p}^{2}d}{\epsilon_{0}c^{2}}\sum_{n}G(\omega_{p},\mathbf{r}_{\perp}-\mathbf{r}^{\perp}_{n},z-z_n)\hat{\sigma}_{ge,n}.
\label{E_r}
\end{equation}
Here, $\hat{E}(\mathbf{r}_{\perp},z)=\mathbf{e}_{d}^{*}\cdot\hat{\mathbf{E}}(\mathbf{r}_{\perp},z)e^{i\omega_pt}$ is the slowly varying envelope of the field around a carrier frequency $\omega_p$ and projected onto the transition-dipole orientation $\mathbf{e}_{d}$, whereas $\hat{E}_0(\mathbf{r}_{\perp},z)$ is the corresponding freely evolving input field [given by Eq. \eqref{SE} with $\hat{a}_{\textbf{k}_\perp k_z\mu}(t)\rightarrow \hat{a}_{\textbf{k}_\perp k_z\mu}(0)e^{-i\omega_{\textbf{k}_\perp k_z}t}$]. $G(\omega,\mathbf{r}_{\perp},z)$ is the (dyadic) Green's function of the photon field in free space at frequency $\omega$, also projected onto the dipole orientation. The total field is then given by a superposition of the incoming field and that emitted by the atoms and propagated via the Green's function.

Under the same Born-Markov approximation, we integrate out the photonic operators and derive the Heisenberg-Langevin equations for the atomic operators. Linearizing the equations in the weak quantum field (e.g. see Supplement of Ref. \cite{Our_paper}), we obtain
\begin{eqnarray}
\frac{d\hat{\sigma}_{ge,n}}{dt}&=&-\left(\frac{\gamma_{s}}{2}-i\delta_{p}\right)\hat{\sigma}_{ge,n}+i\Omega\hat{\sigma}_{gs,n}+\frac{id}{\hbar}\hat{E}_{0}(\mathbf{r}^{\perp}_{n},z_n)\nonumber\\&+&\hat{F}_n+\frac{i}{\hbar}\frac{d^2\omega_{p}^{2}}{\epsilon_{0}c^{2}}\sum_{m}G(\omega_{p},\mathbf{r}^{\perp}_n-\mathbf{r}^{\perp}_{m},z_n-z_m)\hat{\sigma}_{ge,m},
\nonumber\\
\label{sigma_ge_real}
\\
\frac{d\hat{\sigma}_{gs,n}}{dt}&=&i\delta_{2}\hat{\sigma}_{gs,n}+i\Omega^{*}(t)\hat{\sigma}_{ge,n}.
\label{sigma_gs_real}
\end{eqnarray}
Here, $\gamma_{s}$ is a spontaneous decay rate due to non-collective processes: it may include a decay $\gamma_{\mathrm{se}}$ from $\ket{e}$ to $\ket{s}$, and a non-collective decay due to other imperfections, e.g. position disorder in the case of atomic arrays (see below). $\hat{F}_n$ is the corresponding quantum Langevin noise operator satisfying $\langle\hat{F}_n\hat{F}_m^\dagger\rangle=\delta_{nm}\gamma_{s}$, $\delta_p =  \omega_p- \omega_e$ is the detuning of the $\ket{g}\leftrightarrow\ket{e}$ transition and $\delta_2=\delta_p-\delta_c$ is the two-photon detuning, with
$\delta_c = \omega_c -(\omega_e-\omega_s)$ being the detuning of the $\ket{s}\leftrightarrow\ket{e}$ transition (Fig. 1c). The Green's function in Eq. \eqref{sigma_ge_real} describes the dipole-dipole interaction between the $\ket{g}\leftrightarrow\ket{e}$ transition dipoles of pairs of atoms $n$ and $m$; such a term is absent between the $\ket{s}\leftrightarrow\ket{e}$ transition dipoles of different atoms since in the linear regime the probability to excite two atoms to states $\ket{s}$ and $\ket{e}$ is low \cite{Our_paper}.

\subsection{The target photon mode}
We consider incident light propagating along $z$ with a transverse mode profile $u(\mathbf{r}_{\perp})$, satisfying the normalization $\int_{-\infty}^{\infty}|u(\mathbf{r}_{\perp})|^{2}d\mathbf{r}_{\perp}=1$. For a paraxial beam, the typical scale of spatial variations of $u(\mathbf{r}_{\perp})$, denoted by $w$, is much larger than the optical wavelength $\lambda=2\pi c/\omega_p$, as in the typical case of a Gaussian beam $u(\mathbf{r}_{\perp})=\sqrt{\frac{2}{\pi w^2}}e^{-(\mathbf{r}_{\perp})^2/w^2}$. We define the field projected to this mode as
\begin{eqnarray}
\hat{E}_u(z)&\equiv&\frac{1}{\sqrt{A_u}}\int^\infty_{-\infty}\hat{E}(\mathbf{r}_{\perp},z)u^*(\mathbf{r}_{\perp})d\mathbf{r}_{\perp}\nonumber\\&=&i\sum_{k_{z}}\sqrt{\frac{\hbar\omega_{k_{z}}}{2\epsilon_{0}LA_u}}\hat{a}_{u k_{z}}e^{i(k_{z}z+\omega_pt)},
\label{transformed_E}
\end{eqnarray}
where $A_u$ is an unimportant area scale associated with the mode, e.g. $A_u=(\pi/2)w^2$. The second equality, appearing as an expansion in 1D propagating waves, is obtained within the paraxial approximation by introducing the 1D continuum of annihilation operators associated with the transverse mode $u(\mathbf{r}_{\perp})$, $\hat{a}_{u k_{z}}=\frac{2\pi}{L}\sum_{\textbf{k}_\perp}\sum_{\mu}\textbf{e}_{d}^{*}\cdot\mathbf{e}_{\textbf{k}_\perp k_{z}\mu}\hat{a}_{\textbf{k}_\perp k_{z}}\tilde{u}^*(\textbf{k}_\perp)$, with $\tilde{u}^*(\textbf{k}_\perp)$ being the Fourier transform of ${u}^*(\mathbf{r}_{\perp})$, and $[\hat{a}_{u k_{z}},\hat{a}_{u k_{z}'}^\dagger]=\delta_{k_z,k_z'}$.

For a paraxial mode $u(\mathbf{r}_{\perp})$ of spatial width $w\gg\lambda$, and considering propagation distances within its Rayleigh range, $z<z_R=\pi w^2/\lambda$, diffraction of the mode is negligible and its propagation along $z$ is expected to be that of an effective plane wave in 1D. Indeed, performing the projection \eqref{transformed_E} on the field equation \eqref{E_r}, we show in Appendix C that for $w\gg\lambda$ and $z<z_R$ one obtains
\begin{equation}
\hat{E}_u^\pm(z)=\hat{E}_{u,0}^\pm(z)+e^{\pm i k_p z}\frac{i d k_p}{2\epsilon_0\sqrt{A_u}}\sum_n  u^{\ast}(\mathbf{r}^{\perp}_{n})e^{\mp i k_p z_n}\hat{\sigma}_{ge,n}.
\label{in_out_transformed1}
\end{equation}
Here $k_p=\omega_p/c=2\pi/\lambda$, whereas $\hat{E}_{u}^+(z)$ and $\hat{E}_{u}^-(z)$ are the right- and left-propagating fields [including only $k_z>0$ or $k_z<0$, respectively, in Eq. \eqref{transformed_E}], sampled at $z>z_{\mathrm{max}}$ and $z<z_{\mathrm{min}}$ respectively ($z_n\in[z_{\mathrm{min}},z_{\mathrm{max}}]$). We identify that $\hat{E}_u^\pm$ are coupled to $\hat{P}_{\pm}= \sum_n  u^{\ast}(\mathbf{r}^{\perp}_{n})e^{\mp i k_p z_n}\tilde{\sigma}_{ge,n}$. However, for a mapping to Eq. (\ref{inout}) of the two-sided 1D model we need a single dipole mode $\hat{P}=\hat{P}_{+}=\hat{P}_{-}$. A simple way to achieve this is by demanding $e^{i k_p z_n}=e^{-i k_p z_n}$, or
\begin{equation}
z_n=(\lambda/2)\times \mathrm{integer} \quad \quad \forall n.
\label{zn}
\end{equation}
\subsection{The collective dipole}
The considerations above establish the natural ``candidate" for the relevant collective dipole given by
\begin{equation}
\hat{P}= \sum_n  u^{\ast}(\mathbf{r}^{\perp}_{n})e^{- i k_p z_n}\hat{\sigma}_{ge,n},
\label{Pp}
\end{equation}
also considering the condition (\ref{zn}). We now turn to write the equation of motion for $\hat{P}$ and identify the conditions under which it can be mapped to that of the 1D model of Eq. (\ref{model}). To this end, we first introduce the eigenvectors $v_{l,n}$ with eigenvalues $D_l$ of the dipole-dipole interaction kernel, $D_{nm}\equiv-\frac{i}{\hbar}\frac{d^2\omega_{p}^{2}}{\epsilon_{0}c^{2}}G(\omega_{p},\mathbf{r}^{\perp}_n-\mathbf{r}^{\perp}_{m},z_n-z_m)$,
\begin{equation}
\sum_m D_{nm}v_{l,m}=D_lv_{l,n}, \quad D_{nm}=\sum_l v_{l,n}v_{l,m} D_l,
\label{vl}
\end{equation}
noting the orthogonality and completeness relations of eigenvectors of the complex symmetric matrix $D_{nm}$, $\sum_n v_{l,n} v_{l',n}=\delta_{ll'}$, $\sum_l v_{l,n}v_{l,m}=\delta_{nm}$ \cite{Manzoni}. Writing Eq. (\ref{sigma_ge_real}) for $\hat{P}$ using (\ref{Pp}) and (\ref{vl}) we find
\begin{eqnarray}
\frac{d\hat{P}}{dt}&=&\left(i\delta_p-\frac{\gamma_s}{2}\right)\hat{P}+i\Omega\hat{S}+i\frac{d}{\hbar}\hat{E}_0+\hat{F}
\nonumber\\
&-&\sum_l D_l \sum_m v_{l,m}\tilde{\sigma}_{ge,m}\sum_n u^{\ast}(\mathbf{r}^{\perp}_{n})e^{- i k_p z_n} v_{l,n},
\label{Pl}
\end{eqnarray}
where $\hat{S}$, $\hat{E}_0$ and $\hat{F}$ are those from Eq. (\ref{sigma_ge_real}) simply transformed via Eq. (\ref{Pp}). The last term describes the coupling of the collective dipole $\hat{P}$ to dipole eigenmodes of the form $\hat{P}_l=\sum_m v_{l,m}\tilde{\sigma}_{ge,m}$.

Coupling of $\hat{P}$ to other dipole modes prevents the mapping to the 1D model equations (\ref{model}) which exhibit a diagonal form for $\hat{P}$. This can be remedied considering either of the following cases:
(i) The target mode is equal to one specific dipole eigenmode $l=0$, i.e. $u^{\ast}(\mathbf{r}^{\perp}_{n})e^{- i k_p z_n}=v_{l=0,n}$; or
(ii) The target mode $u^{\ast}(\mathbf{r}^{\perp}_{n})e^{- i k_p z_n}$ overlaps only with dipole eigenmodes $l$ which are all quasi-degenerate, i.e. their eigenvalues all satisfy $D_l\approx D_0$.
In both cases the last term becomes $\propto D_0\hat{P}$ and Eq. (\ref{Pl}) appears as that of the 1D model, Eq. (\ref{model}).

The term $\hat{E}_0=\sum_n  u^{\ast}(\mathbf{r}^{\perp}_{n})e^{- i k_p z_n}\hat{E}_0(\mathbf{r}_{\perp},z_n)$ contains the overlap with the target mode but possibly also with other photon modes (see examples below). This means that $\mathrm{Re}[D_0]$ contains contributions both due to emission to the target mode $\Gamma$ and to lossy modes $\gamma_{\mathrm{loss}}$.

\subsection{Conclusion: Conditions for mapping, and the case of atom arrays}
Considering a paraxial target mode we find that the mapping to the 1D model can be established if the target mode profile spatially matches either a single dipole eigenmode or a collection of degenerate dipole eigenmodes. This is in addition to the sufficient condition (\ref{zn}) for the symmetric, two-sided case.

These conditions are natural to satisfy for atomic arrays, which explains why the reflectivity in atom-array interfaces should characterize their operation. For example, consider first an infinite 2D array ($z_n=0$), wherein the lattice translation invariance imposes that the dipole eigenmodes are 2D lattice Fourier modes, $l\rightarrow \mathbf{k}_{\perp}$. Then, a paraxial target mode with a very large waist $w\gg \lambda$ spatially overlaps with dipole modes of a very small wavevector  $|\mathbf{k}_{\perp}| < 2\pi/w \rightarrow 0$. The latter can become quasi-degenerate for the relevant time-scales, $D_{\mathbf{k}_{\bot}}\approx D_{\mathbf{k}_{\bot}=0}$ \cite{Efi4}, thus allowing the mapping to the 1D model. Considering finite-size arrays, this reasoning still holds if the array is large enough, as we show in the next sections. This allows us to establish the mapping to the 1D model also including finite-size effects and imperfections.

\section{Example 1: 2D atom arrays}
We begin with the example of a 2D array in free space (Fig. 2). Most of the discussion is dedicated to arrays whose lattice spacing $a$ are smaller than the relevant optical wavelength $\lambda=2\pi c/\omega_p$, as typical of an optical lattice realization \cite{Bloch1,Bloch2,Rui}, whereas towards the end we also briefly comment on the case $a>\lambda$, relevant to optical tweezer arrays \cite{tweezer1,tweezer2}. Readers not interested in the details of the mapping may skip directly to Sec. V C, where the analysis of the results from Tables I and II is discussed.
\begin{figure}[t!]
  \begin{center}
    \includegraphics[width=\columnwidth]{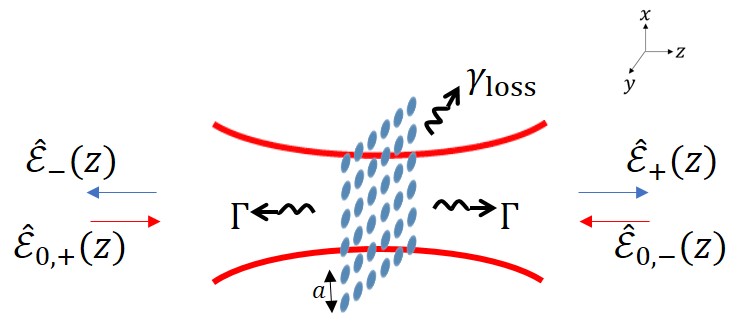}
    \caption{\small{
2D array quantum interface. A finite-size array with lattice spacing $a$ is situated on the $xy$ plane ($z=0$). The target-mode input consists of the symmetric combination of fields from both sides, $\hat{\mathcal{E}}_{0,\pm}$. The full free-space multimode problem is mapped to the generic model of Fig. \ref{fig1}b, with the relevant field and collective dipole modes projected to the target mode-profile [Eqs. (\ref{transformed_E}), (\ref{transform_P}), (\ref{transform_S})] and with the effective parameters $\Gamma$ and $\gamma_{\mathrm{loss}}$ from Table \ref{table:1}.
    }} \label{2D_array}
  \end{center}
\end{figure}
\subsection{System}
We consider a 2D array of identical three-level atoms (Fig. 1c) $n=1,...,N$, forming a square lattice on the $xy$ plane with a lattice spacing $a$ (Fig. \ref{2D_array}). The linear size  of the array is $L_a=a\sqrt{N}$ and subwavelength spacing, $a<\lambda$, is assumed (see Sec. V C for $a>\lambda$). The Hamiltonian and equations of motion are identical to those from Eqs. (\ref{H_s})-(\ref{H_I}) and (\ref{E_r})-(\ref{sigma_gs_real}), with the atomic positions now given by 2D array lattice points,
$z_n=0$ and $\mathbf{r}^{\perp}_n=\mathbf{r}^{\perp}_{n_x,n_y}=a(n_x,n_y)$ ($n_x,n_y$ integers).
Imperfections due to small spatial disorder $\delta r$ in the array atomic positions are accounted for by supplementing the non-collective decay term $\gamma_s$ in Eq. (\ref{sigma_ge_real}) with the scattering rate $\gamma_{\mathrm{dis}}\propto (\delta r/\lambda)^2$ \cite{Efi5,Efi4}, leading to $\gamma_s=\gamma_{\mathrm{se}}+\gamma_{\mathrm{dis}}$ (see further discussion in Sec. V C). We note that the decay rate $\gamma_{\mathrm{se}}$ of the $|s\rangle\leftrightarrow |e\rangle$ transition may become negligible, e.g. in a ladder-type atomic configuration (Fig. \ref{fig1}d) where $\ket{s}$ is a highly excited metastable level (Rydberg state).

\subsection{Mapping to the 1D model}
The array is illuminated from both sides by a paraxial target mode with a normalized transverse profile $u(\mathbf{r}_{\bot})$, as defined in Eq. (\ref{transformed_E}). A typical example is a Gaussian beam $u(\mathbf{r}_{\perp})=\sqrt{\frac{2}{\pi w^2}}e^{-(\mathbf{r}_{\perp})^2/w^2}$, with $w\gg \lambda$. Following the reasoning presented in Sec. IV, we then anticipate that the relevant collective dipoles should follow the same spatial structure. We then define
\begin{eqnarray}
\hat{P}_u&=&\frac{a}{\sqrt{\eta}}\sum_{n}u^*(\mathbf{r}^{\perp}_n)\hat{\sigma}_{ge,n},
\label{transform_P}
\\
\hat{S}_u&=&\frac{a}{\sqrt{\eta}}\sum_{n}u^*(\mathbf{r}^{\perp}_n)\hat{\sigma}_{gs,n},
\label{transform_S}
\end{eqnarray}
where
\begin{equation}
\eta=\int_{L_a^2}|u|^2(\mathbf{r}_{\perp})d\mathbf{r}_{\perp}\quad\xrightarrow[\mathrm{Gaussian}]{}\quad\eta=\mathrm{erf}^2(\frac{L_a}{\sqrt{2}w}),
\label{eta}
\end{equation}
is the fraction of the spatial mode that overlaps with the atomic array of size $L_a^2=a^2N$; in the limit of an array much larger than the beam size $L_a\gg w$, $\eta\rightarrow1$. This normalized definition guarantees the bosonic commutation relation $[\hat{P}_u,\hat{P}_u^\dagger]=[\hat{S}_u,\hat{S}_u^\dagger]=1$ within the linear regime taken here (also assuming $w\gg a$, so that $\sum_{n}\rightarrow \frac{1}{a^2}\int_{L_a^2}d\mathbf{r}_{\perp}$).

\emph{Field equations}.--- Performing the projection \eqref{transformed_E} on the field equation \eqref{E_r}, we obtained Eq. (\ref{in_out_transformed1}) which in terms of $\hat{P}_u$ becomes
\begin{equation}
\begin{split}
&\mathcal{\hat{E}}_u^\pm(z)=\mathcal{\hat{E}}_{u,0}^\pm(z)+i\sqrt{\frac{\eta\Gamma_0}{2}}e^{\pm ik_pz}\hat{P}_u,\\
&\mathcal{\hat{E}}_u^\pm(z)\equiv\frac{d}{\hbar}\sqrt{\frac{2A_u}{a^2\Gamma_0}}\hat{E}_u^{\pm}(z),
\end{split}
\label{in_out_transformed}
\end{equation}
recalling the assumptions $w\gg\lambda$ and $z<z_R$, and with $k_p=\omega_p/c$ and $\hat{E}_{u}^{\pm}(z)$ being the right- and left-propagating fields (see Appendix D). The effective 1D coupling strength to $\hat{P}_u$ is seen to be given by $\Gamma_0$, which is the collective emission rate of a uniformly excited infinite array, and given by \cite{Efi5}
\begin{equation}
\Gamma_0=\frac{3}{4\pi}\frac{\lambda^2}{a^2}\gamma,\quad\gamma=\frac{d^2\omega_p^3}{3\pi\epsilon_0\hbar c^3},
\end{equation}
with $\gamma$ being the usual free-space spontaneous emission rate of a single atom  ($\ket{g}\leftrightarrow\ket{e}$ transition).

The scattered field propagates symmetrically on both sides of the array, so we define the total symmetric field $\mathcal{\hat{E}}_u(z)$ as a symmetric superposition of the right and left propagating fields $\mathcal{\hat{E}}_u(z)=\frac{1}{\sqrt{2}}\bigg[\mathcal{\hat{E}}_u^+(z)+\mathcal{\hat{E}}_u^-(-z)\bigg]e^{-ik_pz}$, ($z>0$). The input-output relation for this symmetric field then reads
\begin{equation}
\mathcal{\hat{E}}_u(z)=\mathcal{\hat{E}}_{u,0}(z)+i\sqrt{\eta\Gamma_0}\hat{P}_u.
\label{in_out_u}
\end{equation}
This equation reveals the mode-preserving light-matter coupling: an input light $\mathcal{\hat{E}}_{u,0}$ at the target mode
 $u(\mathbf{r}_{\perp})$ excites the corresponding collective dipole mode $\hat{P}_u$ of the atomic array, which will finally be emitted as an output light $\mathcal{\hat{E}}_{u}$ of the same target mode. The effective coupling strength is $\eta\Gamma_0$, where the geometrical-overlap factor $\eta$ can be understood as a correction to the ideal coupling strength $\Gamma_0$. It accounts for the fact that a fraction $(1-\eta)$ of the beam extends beyond the finite-size array and hence does not interact with the atoms.

\emph{Atom equations}.--- We now show how the dipole mode $\hat{P}_u$ that matches $\mathcal{\hat{E}}_{u,0}$, approximately diagonalizes the dipole-dipole kernel in Eq. (\ref{sigma_ge_real}). Transforming Eq. (\ref{sigma_ge_real}) according to \eqref{transform_P}, we assume the following: (i) $L_a\gg\lambda$, the array length is larger than the effective range of the dipole-dipole interaction, so each atom in the "bulk" (not at the edges) effectively feels interactions of an infinite array. (ii) $\sqrt{N}\gg1$, so most atoms are in the bulk, and edge atoms are negligible in describing collective dipoles. With these two assumptions we expect that the finite size of the array is manifested by its non-perfect overlap with the beam, while collective dipole-dipole shifts and widths are well approximated by those of a uniformly excited infinite array, $\Delta_{0}$ and $\Gamma_{0}$, respectively. Indeed, the full derivation in Appendix D yields
\begin{eqnarray}
\frac{d\hat{P}_u}{dt}&=&-\left[\frac{\Gamma_0}{2}+\frac{\gamma_s}{2}+i\left(\Delta_{0}-\delta_{p}\right)\right]\hat{P}_u+i\Omega\hat{S}_u\nonumber\\&+&i\sqrt{\eta\Gamma_0}\mathcal{\hat{E}}_{u,0}(0)+\hat{F}_u+\hat{F}_\eta,
\label{P_1}
\end{eqnarray}
\begin{equation}
\frac{d\hat{S}_u}{dt}=i\delta_{2}\hat{S}_u+i\Omega^{*}(t)\hat{P}_u,
\label{S}
\end{equation}
with the Langevin noises $\langle \hat{F}_u(t)\hat{F}_u^\dagger(t')\rangle=\gamma_s\delta(t-t')$ and $\langle \hat{F}_\eta(t)\hat{F}^\dagger_\eta(t')\rangle\approx(1-\eta)\Gamma_0\delta(t-t')$. While $\hat{P}_u$ exhibits a collective decay $\Gamma_0$, we observe that its coupling to the target mode is characterized by $\eta\Gamma_0$, expressing the fraction $\eta<1$ emitted to it. We therefore divide $\Gamma_0$ in Eq. \eqref{P_1} to $\Gamma=\eta\Gamma_0$ and $(1-\eta)\Gamma_0$, where the latter expresses the fraction of the collective component of the emission which is not coupled to the target mode, as discussed below (see also Fig. \ref{losses}b). Therefore, the total emission outside of the target mode is given by the sum of the latter and the non-collective emission $\gamma_s$, yielding
\begin{eqnarray}
\frac{d\hat{P}_u}{dt}&=&-\left[\frac{\eta\Gamma_0}{2}+\frac{\gamma_{\mathrm{loss}}}{2}+i\left(\Delta_{0}-\delta_{p}\right)\right]\hat{P}_u+i\Omega\hat{S}_u\nonumber\\&+&i\sqrt{\eta\Gamma_0}\mathcal{\hat{E}}_{u,0}(0)+\hat{F},
\label{P}
\end{eqnarray}
with
\begin{equation}
\gamma_{\mathrm{loss}}=(1-\eta)\Gamma_{0}+\gamma_{s}.
\label{gamma_loss}
\end{equation}
Correspondingly the Langevin noise $\hat{F}=\hat{F}_u+\hat{F}_\eta$ is comprised of Langevin noises due to these two loss effects and it satisfies $\langle\hat{F}(t)\hat{F}^\dagger(t')\rangle=\gamma_{\mathrm{loss}}\delta(t-t')$. For an array sufficiently larger than the target mode waist $w$, we have $\eta=1$, and the collective emission is fully directed to the target mode.

It is easy to see that the dynamical equations of the atomic array, Eqs. \eqref{S}-\eqref{P}, together with the input-output relation Eq. \eqref{in_out_u} are completely equivalent to the dynamical equations of the 1D model Eq. \eqref{model2}, with the effective parameters $\Gamma=\eta\Gamma_0$ and $\gamma_{\mathrm{loss}}$ from Eq. \eqref{gamma_loss}, as also summarized in Table \ref{table:1}. This holds for any target paraxial mode $u(\mathbf{r}_{\perp})$ within its Rayleigh range.

\subsection{Cooperativity and efficiency analysis}
As explained above, once the mapping to the 1D model of Eqs. \eqref{model2} is established, the efficiency of the light-matter interface is completely determined by the effective parameters $\Gamma$ and $\gamma_{\mathrm{loss}}$ (Table \ref{table:1}), and the resulting cooperativity $C=\Gamma/\gamma_{\mathrm{loss}}$ or resonant reflectivity $r_0=C/(C+1)$.

Beginning with the desired emission to the target mode, $\Gamma=\eta\Gamma_0$, it is given by the effective 1D emission rate multiplied by the overlap factor $\eta$ between the array and target-mode cross-sections, Eq. \eqref{eta}.

\begin{figure}[h!]
  \begin{center}
    \includegraphics[width=\columnwidth]{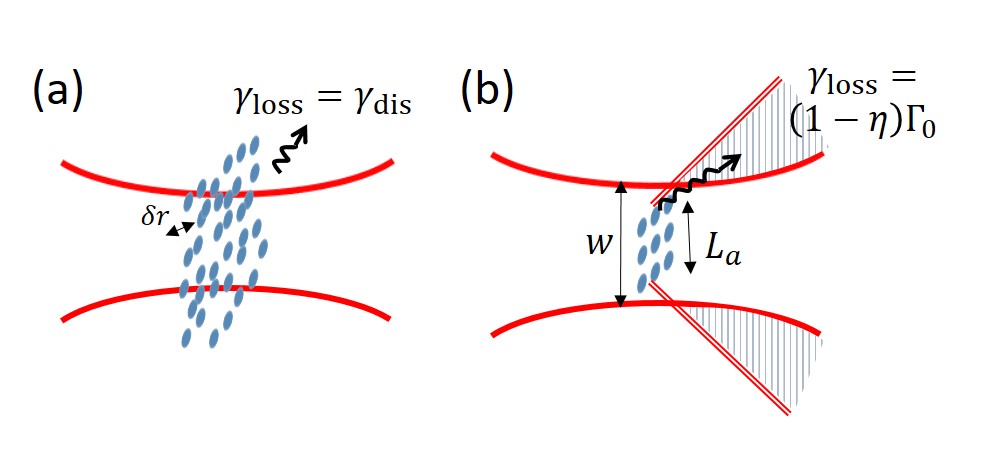}
    \caption{\small{
Sources of imperfection and loss in an atom array. (a) Disorder $\delta r$ in atomic positions around the perfect lattice points leads to an effectively individual-atom scattering to undesired modes at a rate $\gamma_{\mathrm{dis}}$. (b) The finite-size of the array (linear size $L_a=\sqrt{N}a$), leads to its non-ideal overlap with the target mode, $\eta=\mathrm{erf}^2(\frac{L_a}{\sqrt{2}w})<1$, and hence to a fraction ($1-\eta$) of the collective emission which is spilled over to undesired modes.
The error scaling induced by these loss mechanisms are summarized in Table \ref{table:2}.
    }} \label{losses}
  \end{center}
\end{figure}

The undesired emission rate $\gamma_{\mathrm{loss}}$ from Eq. \eqref{gamma_loss} has two contributions. The individual-atom rate $\gamma_s=\gamma_{\mathrm{se}}+\gamma_{\mathrm{dis}}$ includes the emission rate from $\ket{e}$ to $\ket{s}$ (recalling it may become negligible in ladder-type atoms, Fig. \ref{fig1}d), and the disorder-induced scattering. For the latter, we consider a standard deviation $\delta r$ in atomic positions around the array lattice positions $\mathbf{r}_n$ at all directions $x,y,z$ (Fig. \ref{losses}a). This breaks translation invariance and causes scattering to directions other than that of the target mode. To lowest order, this causes an effective individual atom scattering rate which can be shown to scale as $\gamma_{\mathrm{dis}}\sim(2\pi \delta r/\lambda)^2\Gamma_0$ \cite{Efi1,Efi3,Efi5}.

For an array comparable or smaller than the target beam cross-section ($L_a\lesssim w$), where $\eta<1$, the collective decay also contributes to the undesired emission the rate $(1-\eta)\Gamma_0$. This can be intuitively understood from Fig. \ref{losses}b as follows: light diffracted from the edges of the array has an angular spread of $~\lambda/L_a$, which could be larger than the angular spread $~\lambda/w$ contained in the target mode, so that a portion $(1-\eta)$ of the light emitted from the array is not emitted into the desired target mode. For a target Gaussian mode, where $\eta$ is given by $\mathrm{erf}^2(\frac{L_a}{\sqrt{2}w})$ [Eq. \eqref{eta}], we obtain the first correction to the ideal $\eta=1$ case by expanding to first order around $L_a/w\rightarrow\infty$, obtaining $1-\eta=\frac{2}{\sqrt{\pi}}\frac{\mathrm{exp}[-L_a^2/(2w^2)]}{L_a/(\sqrt{2}w)}$. For a quantum memory, this error can then scale exponentially better with the size of the array, or equivalently with the number of atoms, as was also found numerically in \cite{Facchinetti,Manzoni} (for 1D arrays, an exponential improvement was found in the context of subradiant modes \cite{Asenjo-Garcia}). It should be noted however, that taking other sources of errors into account, such as the dispersion of the Gaussian beam neglected here, the exponential improvement is not necessarily observed \cite{Manzoni}. However, for sufficiently wide Gaussian beams of even a few $\lambda$ , this beam dispersion becomes negligible and the exponential scaling dominates as also seen below.

\begin{table}[h!]
\centering
\begin{tabular}{| >{\centering\arraybackslash}m{7em} | >{\centering\arraybackslash}m{7em} | >{\centering\arraybackslash}m{11em}|}
\hline
Loss mechanism & Effective error & Scaling \\
\hline\hline
Disorder & $\frac{1}{C}=\frac{\gamma_{\mathrm{dis}}}{\Gamma_0}$ & $(2\pi \delta r/\lambda)^2$ \\ [2ex]
\hline
Finite size array & $\frac{1}{C}=\frac{(1-\eta)\Gamma_0}{\eta\Gamma_0}$ & $\frac{2}{\sqrt{\pi}}\frac{\sqrt{2}w}{L_a}\mathrm{exp}[-L_a^2/(2w^2)]$ \\ [2ex]
\hline
\end{tabular}
\caption{
Scaling of the error $1-r_0=1/(C+1)\approx1/C$ due to different sources of imperfection and loss. The error due to position disorder (Fig. \ref{losses}a) scales quadratically with the standard deviation $\delta r$ about the perfectly ordered array positions \cite{Efi3,Efi5,Efi6}. The error due to the imperfect overlap between the target mode and the array due to the latter's finite size (Fig. \ref{losses}b) scales essentially exponentially better with the atom number $N=L_a^2/a^2$ (asymptotic result valid for $L_a\gg w$, see text).
}
\label{table:2}
\end{table}

The scaling of the undesired rates due to both imperfections $\gamma_{\mathrm{dis}}$ and $(1-\eta)$ are summarized in Table \ref{table:2}. To test these analytical predictions, we can use an important principle revealed in this work, i.e. that the cooperativity $C$ can be extracted from a simple classical calculation of the reflectivity. To this end, we perform classical numerical calculations of the scattering of a Gaussian beam off an array in different configurations \cite{Efi5}. We scan the frequency of the incidence field looking for peaks in the intensity of the reflected field, thus finding the on-resonance reflectivity $R=r_0^2=\frac{C^2}{(1+C)^2}$ from which we extract the effective loss $1/C=\gamma_{\mathrm{loss}}/\Gamma$ (or cooperativity $C$). We note that this formalism treats each atom as a two-level system and does not include the possible individual decay $\gamma_{\mathrm{se}}$ due to the $\ket{e}\rightarrow\ket{s}$ transition, so the only sources of loss are the non-collective emission due to the disorder $\gamma_s=\gamma_{\mathrm{dis}}$ and the  imperfect overlap of the beam with the array $(1-\eta)\Gamma_0$.

\begin{figure}[t!]
  \begin{center}
    \includegraphics[width=\columnwidth]{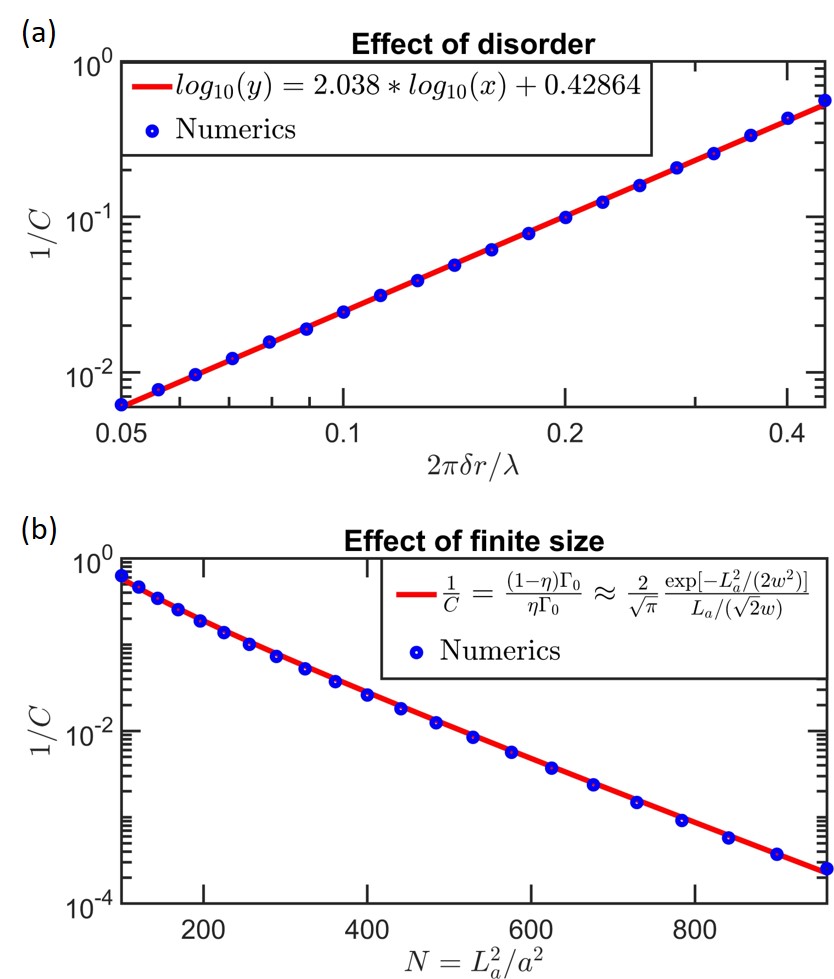}
    \caption{\small{
Cooperativity $C$ extracted from classical numerical calculations of the reflectivity in 2D arrays (Sec. V C). Agreement with the predicted analytical scaling of Table \ref{table:2} is exhibited, in correspondence to the dominant source of imperfection. (a) Disorder imperfection: $1/C$ plotted as a function of the standard deviation $\delta r$ around the perfectly ordered array positions. The linear fit to the log-log plot verifies the scaling $\gamma_{\mathrm{dis}}/\Gamma_0\sim(2\pi\delta r/\lambda)^2$ predicted analytically for a large enough array with respect to the incident-beam waist, $L_a=\sqrt{N}a>w$, as taken here (Table \ref{table:2} and text). Numerical parameters: lattice spacing $a/\lambda=0.6$, atom number $N=30\times30$, waist $w=0.25 L_a$. Scattered fields were calculated at a distance $z=5\lambda$ from the array and projected to the Gaussian beam to extract the resonant reflectivity $r_0$ and $C$ (averaged over $50$ disorder realizations).
(b) Effect of finite array-size: $1/C$ plotted as a function of array atom number $N=L_a^2/a^2$ for a perfectly-ordered array. The error $1/C$ is seen to scale exponentially better with $N$, as predicted analytically (Table \ref{table:2} and text) and originating in the error-function behavior of the overlap $\eta$ between the array and the Gaussian target mode.
Here the waist was fixed to $w/a=8$, while $a/\lambda=0.6$ as before.
    }} \label{Sub_numerics}
  \end{center}
\end{figure}

We begin by considering weak disorder in array positions for an array size sufficiently larger than the beam waist. The latter means that effectively $\eta=1$ and $\gamma_{\mathrm{loss}}$ is completely dominated by the disorder, i.e. $\frac{1}{C}=\frac{\gamma_{\mathrm{dis}}}{\Gamma_0}$. The results for $1/C$, extracted  numerically as described above, are plotted in Fig. \ref{Sub_numerics}a as a function of the standard deviation of atomic positions. The linear fit to the log-log plot confirms that $\gamma_{\mathrm{dis}}$ scales quadratically with the standard deviation of the atomic positions $\gamma_{\mathrm{dis}}\sim(2\pi\delta r/\lambda)^2\Gamma_0$, as predicted analytically (Table \ref{table:2}). Considering now the effect of a finite-size array, Fig. \ref{Sub_numerics}b shows the numerical calculation of $1/C$ as a function of the number of atoms $N=L_a^2/a^2$ for a fixed beam waist. In this case the atoms are perfectly ordered, so the only unwanted emission rate is due to the imperfect overlap between the array and the beam, $\frac{1}{C}=\frac{(1-\eta)\Gamma_0}{\eta\Gamma_0}$. It is seen that the numerical calculation agrees very well with this analytical expression for a Gaussian beam, $\eta=\mathrm{erf}^2(\frac{L_a}{\sqrt{2}w})$. Exponential scaling is observed within the parameter regime of our calculations, in agreement with the approximated expression from Table \ref{table:2}.

Finally, we comment on the situation where the lattice spacing $a$ may exceed the wavelength $\lambda$. In this case, even for an infinite array and plane wave illumination ($w\rightarrow \infty$), scattering of light from the collective dipole $\hat{P}_u$ exists in multiple diffraction orders and not only in the zeroth-order, incident field direction \cite{Efi5}. These additional diffraction orders lead to additional loss channels with a total emission rate $\gamma_{\mathrm{diff}}$ which is added to $\gamma_{\mathrm{loss}}$ (Table I), and given by \cite{Efi5,ofernir}
\begin{eqnarray}
\gamma_{\mathrm{diff}}=\Gamma_0\sum_{(m_x,m_y)\in\mathrm{LC},\neq(0,0)}\frac{1-\frac{\lambda^2}{a^2}|(m_x,m_y)\cdot\textbf{e}_d|^2}{\sqrt{1-\frac{\lambda^2}{a^2}(m_x^2+m_y^2)}}.
\label{diff}
\end{eqnarray}
Here the sum is taken over all diffraction orders $(m_x,m_y)\neq (0,0)$ (with $m_x,m_y$ integers) that are within the light-cone $\mathrm{LC}$, i.e. which satisfy ${|(m_x,m_y)|<a/\lambda}$ and describe propagating waves.

\section{Example 2: 3D atom arrays}
We proceed by extending our formalism to include 3D arrays consisting of multiple layers of the 2D array discussed so far. We show that by correctly choosing the layers separation, the cooperativity can increase linearly with the number of layers.
\subsection{System: propagation between 2D arrays}
\begin{figure}[t!]
  \begin{center}
    \includegraphics[width=0.75\columnwidth]{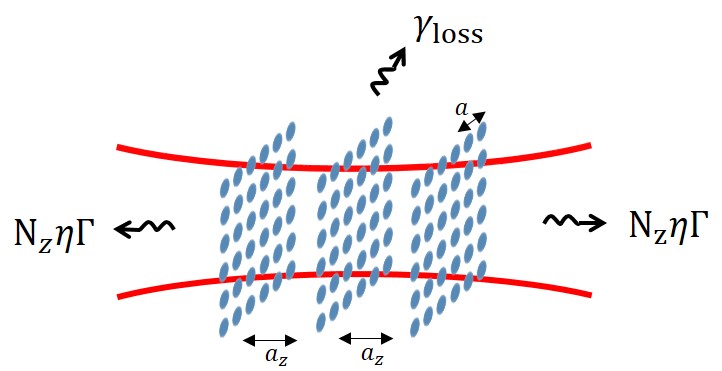}
    \caption{\small{
A 3D atomic array is described as consisting of $N_z$ layers, each being a 2D array (here $N_z=3$). For an inter-layer distance $a_z$ which is an integer multiple of $\lambda/2$, phase-matching occurs, and the cooperativity essentially increases linearly with $N_z$ [see Table \ref{table:1} or Eq. \eqref{P_3D_final}].
    }} \label{3D_array}
  \end{center}
\end{figure}

We consider a 3D array consisting of $N_z$ layers of the 2D array discussed earlier, as illustrated in figure \ref{3D_array}. Each 2D layer consists of $N$ atoms such that the total number of atoms is $N\times N_z$. The layers are positioned at $z=0,a_z,2a_z...,(N_z-1)a_z$, with longitudinal lattice spacing $a_z$ not necessarily equal to the lattice constant $a$ in the $xy$ plane. We assume that the paraxial approximation holds, and that all the layers are within the Rayleigh ranges defined by the finite sizes of both the beam and the layers, $z_R=\pi w^2/\lambda$ and $\sim L_a^2/\lambda$, respectively. The latter condition leads to an approximately diffractionless quasi-1D propagation between the layers, as we see below.

The index $n$ of atomic positions is now comprised of an intra-layer $xy$ index $n_{\bot}=(n_x,n_y)$ and the index $n_z$ of different layers, yielding the lattice positions $(\mathbf{r}^{\bot}_{n};z_n)=(a n_x,a n_y; a_z n_z)$ and corresponding atomic operators $\hat{\sigma}_{\alpha\alpha',n_{\bot},n_z}$. The dynamics of the collective dipole of the layer $n_z$ defined by
\begin{equation}
\hat{P}_{u,n_z}=\frac{a}{\sqrt{\eta}}\sum_{n_{\bot}}u(\mathbf{r}^{\perp}_n)\hat{\sigma}_{ge,n_{\bot},n_z},
\label{transform_P_3D}
\end{equation}
are given by (see Appendix E)
\begin{eqnarray}
\frac{d\hat{P}_{u,n_z}}{dt}&=&-\left[\frac{\Gamma_0}{2}+\frac{\gamma_s}{2}+i(\Delta_0-\delta_{p})\right]\hat{P}_{u,n_z}\nonumber\\&+&i\Omega\hat{S}_{u,n_z}+i\sqrt{\eta\Gamma_0}\mathcal{\hat{E}}_{u,0,n_z}\nonumber\\&+&\hat{F}_{n_z}-\sum_{m_z\neq n_z}D_{n_z,m_z}\hat{P}_{u,m_z},
\label{P_nz}
\end{eqnarray}
with $\mathcal{\hat{E}}_{u,0,n_z}=\frac{d}{\hbar}\frac{\sqrt{A_u}}{a\sqrt{\Gamma_0}}\hat{E}_{u,0}(a_zn_z)$.
Eq. (\ref{P_nz}) describes the coupling between the collective dipoles $\hat{P}_{u,n_z}$ of the layers via the effective inter-layer dipole-dipole kernel
\begin{eqnarray}
D_{n_z,m_z}&=&\frac{\Gamma_0}{2}\sum_{(m_x,m_y)}\frac{1-\frac{\lambda^2}{a^2}|(m_x,m_y)\cdot\textbf{e}_d|^2}{\sqrt{1-\frac{\lambda^2}{a^2}(m_x^2+m_y^2)}}\times\nonumber\\&&e^{ik_p\sqrt{1-\frac{\lambda^2}{a^2}(m_x^2+m_y^2)}a_z|n_z-m_z|}.
\label{D}
\end{eqnarray}
The sum over $(m_x,m_y)$ accounts for contributions from the diffraction orders of each 2D layer. It is seen that every diffraction order $(m_x,m_y)$ mediates a quasi-1D interaction between the layers, $\sim e^{ik_z^{m_x,m_y}a_z|n_z-m_z|)}$, with $k_z^{m_x,m_y}=k_p\sqrt{1-(\lambda^2/a^2)(m_x^2+m_y^2)}$. The latter can be either long-range or evanescent $(k_z^{m_x,m_y}$ real or imaginary, respectively), depending on the lattice spacing $a/\lambda$ and the order $(m_x,m_y)$, similar to the interaction mediated by a multimode waveguide \cite{Efi7}. Considering first a subwavelength array, $a<\lambda$, only $(m_x,m_y)=(0,0)$ can propagate, obtaining
\begin{equation}
\begin{split}
&D_{n_z,m_z}=\frac{\Gamma_0}{2}e^{ik_pa_z|n_z-m_z|}+i\varepsilon_{n_z,m_z},
\end{split}
\label{mixing_term}
\end{equation}
with
\begin{eqnarray}
\varepsilon_{n_z,m_z}&=&\frac{\Gamma_0}{2}\sum_{(m_x,m_y)\neq(0,0)}\frac{\frac{\lambda^2}{a^2}|(m_x,m_y)\cdot\textbf{e}_d|^2-1}{\sqrt{\frac{\lambda^2}{a^2}(m_x^2+m_y^2)-1}}\times\nonumber\\&&e^{-k_pa_z|n_z-m_z|\sqrt{\frac{\lambda^2}{a^2}(m_x^2+m_y^2)-1}}.
\label{epsilon}
\end{eqnarray}
The first term is the Green's function of electrodynamics in 1D, leading to 1D dipole-dipole interaction familiar from 1D ``waveguide" QED \cite{1d_qed,coop_Chang2,Efi7}: Within this picture each 2D array forms a dipole in 1D, and these dipoles exhibit a collective emission $\mathrm{cos}(k_pa_z|n_z-m_z|)$ and frequency shift $\mathrm{sin}(k_pa_z|n_z-m_z|)$. The second term $\varepsilon_{n_z,m_z}$ is a deviation from the purely 1D picture, as it describes the interaction via the evanescent waves of higher diffraction orders. Their evanescent character dictates an exponentially decaying interaction at a range of $\xi_{m_x,m_y}=a/(2\pi\sqrt{m_x^2+m_y^2-(a/\lambda)^2})$, which becomes shorter for increasing orders $m_x,m_y$ and with a typical lengthscale $a$. Assuming an inter-layer distance $a_z\gtrsim\xi_{1,0}$, we can thus treat the term $\varepsilon_{n_z,m_z}$ as a perturbation.

\subsection{Mapping to the 1D model}
First, we need to define a collective dipole $\hat{P}$ of the atoms, that will approximately diagonalize the atomic equations of motion. Here, this amounts to diagonalizing the effective dipole-dipole kernel between layers from Eq. \eqref{mixing_term}. Treating the second term, $\varepsilon_{n_z,m_z}$, as a perturbation as explained above, we consider eigenmodes of the 1D dipole-dipole kernel $e^{ik_pa_z|n_z-m_z|}$. A natural choice is a phase-matched dipole of the form,
\begin{equation}
\hat{P}_u=\frac{1}{\sqrt{N_z}}\sum_{n_z=0}^{N_z-1}\hat{P}_{u,n_z}e^{-ik_pa_zn_z}, \quad 2a_z/\lambda\in\mathbb{N},
\label{P_u}
\end{equation}
with a similar definition for $\hat{S}_u$. Note that we now fix $a_z$ to be an integer multiple of half a wavelength, and that this choice corresponds to $z_n=a_z n_z=(\lambda/2)\times (\mathrm{integer})$, as in our general mapping condition (\ref{zn}). For $2a_z/\lambda$ even, this yields a collective dipole where all layers are in phase, whereas for $2a_z/\lambda$ odd, the layers exhibit alternating signs. Applying this transformation to Eq. \eqref{P_nz}, we obtain (Appendix E):
\begin{eqnarray}
\frac{d\hat{P}_{u}}{dt}&=&-\left[\frac{\eta N_z\Gamma_0}{2}+\frac{\gamma_{\mathrm{loss}}}{2}+i(\Delta_0+\Delta'-\delta_{p})\right]\hat{P}_{u}\nonumber\\&+&i\Omega\hat{S}_{u}+i\sqrt{\eta\Gamma_0N_z}\mathcal{\hat{E}}_{u,0}(0)+\hat{F},
\label{P_3D_final}
\end{eqnarray}
with
\begin{equation}
\gamma_{\mathrm{loss}}=(1-\eta)N_z\Gamma_0+\gamma_s.
\end{equation}
Here, we used the fact that under the Markov approximation  and for $2a_z/\lambda\in\mathbb{N}$, the input field is identical for all layers (up to an alternating signs in the case of $2a_z/\lambda$ odd) $\mathcal{\hat{E}}_{u,0,n_z}=\mathcal{\hat{E}}_{u,0}(0)e^{ik_pa_zn_z}$. This result exhibits several features. First, we observe that the choice of the phase-matched dipole under the condition $2a_z/\lambda\in\mathbb{N}$, yields a collective decay to the target mode which increases linearly with the number of layers $N_z$, in analogy to optical depth in dilute atomic ensembles. Second, the collective component in $\gamma_{\mathrm{loss}}$ also gains a factor $N_z$. This is due to the approximation $z<L_a^2/\lambda$, where diffraction of the losses from the edges is negligible, such that these losses add up coherently as well. Finally, the additional frequency shift $\Delta'$ comes from the perturbative part of the effective inter-layer kernel, $\varepsilon_{n_z,m_z}$, from Eq. \eqref{mixing_term}. Within first order perturbation theory, its effect can be estimated as the matrix element between the collective dipole mode, $\Delta'=\frac{1}{{N_z}}\sum_{n_z=0}^{N_z-1}\sum_{m_z\neq n_z}^{N-1}\varepsilon_{n_z,m_z}e^{ik_pa_z(n_z-m_z)}$. In the regime under study, where $a_z\geq\lambda/2$, it is sufficient to consider only these first order corrections as verified in Appendix F.

The analogous equation for the photon field is found to be (Appendix E)
\begin{equation}
\mathcal{\hat{E}}_u(z)=\mathcal{\hat{E}}_{u,0}(z)+i\sqrt{\eta N_z\Gamma_0}\hat{P}_u,
\label{E_3D}
\end{equation}
whereas the equation for $\hat{S}_u$ is trivially obtained in the usual form of Eq. \eqref{model2}.

The set of equations we obtained are the same as for a single layer with the replacement $\Gamma_0\rightarrow N_z\Gamma_0$. Indeed, the choice of a phase-matched collective operator enhances the emission to the target mode by the number of layers $N_z$. If the correction due to the mismatch in the overlap between the array and the beam is small, $\eta\approx1$, then the cooperativity is also enhanced linearly by the number of the layers $C=\frac{\eta N_z\Gamma_0}{\gamma_{\mathrm{loss}}}$. We note that although this analysis assumed layers comprised of subwavelength 2D arrays ($a<\lambda$), the $N_z$ enhancement of the cooperativity may also apply for $a>\lambda$: this holds if the inter-layer distance $a_z$ is large enough so that the higher diffraction orders are scattered outside and not between the layers.

\section{Example 3: Quantum memory using collective subradiant states}
An atomic quantum memory stores information in a stable collective spin $\hat{S}$ as in the generic model of Sec. III D. Typically, a pair of stable individual-atom states $|g\rangle$ and $|s\rangle$ of a three-state atom are identified, and their collective spin forms $\hat{S}$. In contrast, a recent interesting proposal harnesses collective radiation effects to achieve a quantum memory even in a 2D array of two-level atoms, where a pair of stable states does not exist at the individual atom level \cite{Yelin}. In this case a collective subradiant dipole, which exhibits a suppressed decay due to a destructive interference of the radiated field, forms the stable spin $\hat{S}$. We now show how a realistic 2D array of two-level atoms can be mapped to the generic model of Eq. \eqref{model2} used for the quantum memory protocol. We find analytically the effective model parameters which are essentially identical to those of a 2D array of three-level atoms (Sec. V): Namely, the reflectivity or cooperativity of a realistic 2D array (Table  \ref{table:1}) again determines the quantum memory efficiency; and this result is independent of whether the memory is realized using a third stable atomic level (Sec. V), or using a collective subradiant mode (here). This generic result explains the similarity between the memory efficiencies found in the 2D array numerical studies of Ref. \cite{Manzoni} (three-level atoms) and Ref. \cite{Yelin} (subradiant), respectively.

\subsection{Subradiant dipole as a collective stable state}
\begin{figure}[t!]
  \begin{center}
    \includegraphics[width=\columnwidth]{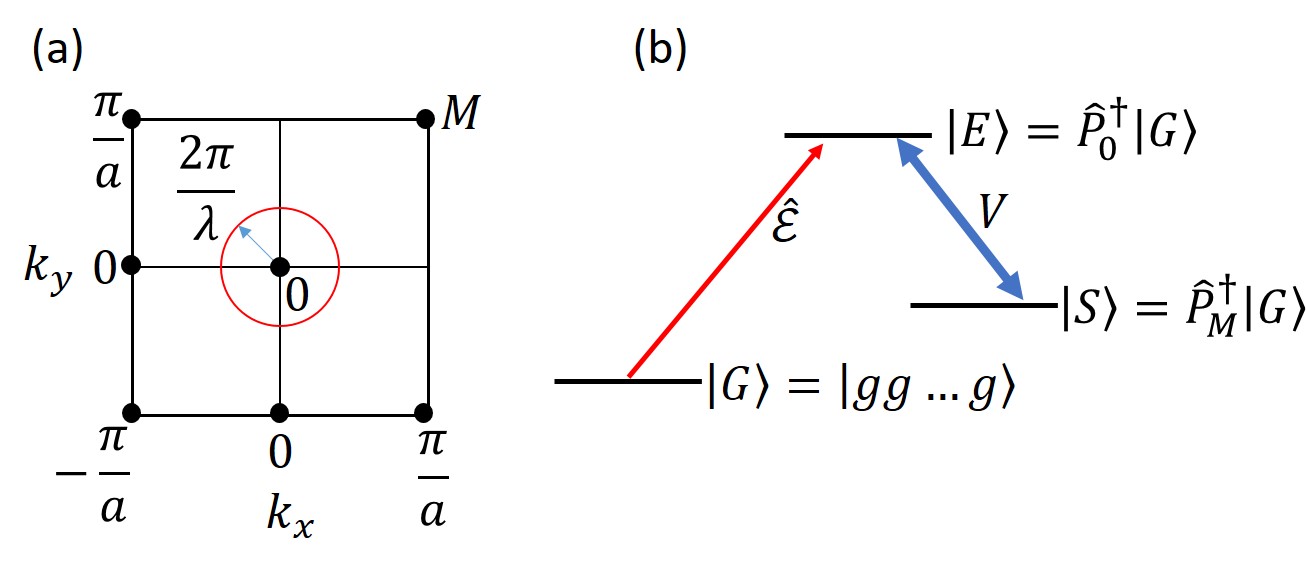}
    \caption{\small{Collective subradiant states of a 2D array. (a) The Brillouin zone of a 2D array with $a<\lambda/\sqrt{2}$ contains regions beyond the ``light-cone" of radius $2\pi/\lambda$. These points, such as the point $M$, describe subradiant collective excitations. (b) Three collective-dipole levels used for the quantum memory scheme of Sec. VII. The level $\ket{S}$ is long-lived since it is formed by a subradiant collective excitation (point $M$). Its coupling to the radiating collective excitation $\ket{E}$ is induced via the spatial energy modulation $V$ (see text).
    }} \label{fig7}
  \end{center}
\end{figure}

To understand the existence of non-radiating subradiant modes, consider first an infinite 2D array of two-level atoms (levels $|g\rangle$ and $|e\rangle$ in Fig. 1c). The collective dipole modes that diagonalize the dipole-dipole kernel are in-plane Fourier modes $\mathbf{k}_{\bot}$ within the first Brillouin zone $k_{x,y}\in\{-\pi/a,\pi/a\}$. Due to lattice symmetry each such mode is coupled to light with the same in-plane momentum and a corresponding longitudinal wavenumber $k_z=\sqrt{(2\pi/\lambda)^2-|\mathbf{k}_{\bot}|^2}$. For lattice spacing satisfying $a<\lambda/\sqrt{2}$ there exist modes for which $k_z$ is imaginary, and correspondingly the collective decay rate is zero, as seen in Fig. \ref{fig7}a \cite{Efi5,Asenjo-Garcia}. In particular, the mode $\mathbf{k}_M=(\pi/a,\pi/a)$ at the corner of the Brillouin zone ($M$-point) is the first to become subradiant, and is chosen in Ref. \cite{Yelin} as the stable spin for quantum memory. The excitation of this subradiant dipole defines a collective stable state $|S\rangle$, whereas the excitation of the symmetric collective dipole ($0$-point in Fig. \ref{fig7}a) defines the collective radiating state $|E\rangle$, which together with the ground state $|G\rangle=|gg...g\rangle$, form a collective version of the typical three-level scheme (Fig. \ref{fig7}b). The necessary coupling between $|E\rangle$ and $|S\rangle$ is achieved via a momentum kick of $\mathbf{k}_M$ provided by a checkerboard-like perturbation of the atomic energy levels as can be realized in an optical superlattice \cite{Li}.

\subsection{Mapping to the 1D model}
We begin with Eqs. \eqref{E_r} and \eqref{sigma_ge_real} for two-level atoms, i.e. where $\Omega=0$ and the level $|s\rangle$ is irrelevant. We add to Eq. \eqref{sigma_ge_real} a local-detuning term, $\delta_p\rightarrow \delta_p+\delta_n$, for an atom $n$ at a 2D lattice point $\mathbf{r}^{\perp}_n=a(n_x,n_y)$, with
\begin{equation}
\delta_n=\delta_{n_x,n_y}=V e^{i\mathbf{k}_M\cdot \mathbf{r}_n}=V(-1)^{n_x+n_y}.
\end{equation}
We now define a pair of collective dipoles corresponding to the points $0$ and $M$ from Fig. \ref{fig7}a,
\begin{equation}
\hat{P}_{0,M}=\frac{a}{\sqrt{\eta}}\sum_n \hat{\sigma}_{ge,n}u(\mathbf{r}^{\perp}_n) e^{i\mathbf{k}_{0,M}\cdot\mathbf{r}^{\perp}_n},
\end{equation}
where $\mathbf{k}_0=0$. $\hat{P}_{0}$ is recognized as the radiating collective dipole from Eq. \eqref{transform_P}, whereas $\hat{P}_M$ is the finite-array analog of the ideal subradiant dipole $M$.
Using Eqs. \eqref{E_r} and \eqref{sigma_ge_real} we obtain coupled equations for these two dipoles and the light, taking the form of Eqs. \eqref{in_out_u}-\eqref{S} with the following replacements:
\begin{equation}
\hat{P}_u\rightarrow \hat{P}_{0}, \quad \hat{S}_u\rightarrow \hat{P}_M, \quad \Omega\rightarrow V, \quad \delta_2\rightarrow \delta_p-\Delta_M+i\frac{\gamma_s}{2}.
\end{equation}
The latter term describes the analog of the two-photon detuning of the stable spin $\hat{S}$ and is derived from Eq. \eqref{sigma_ge_real} transformed to $\hat{P}_M$ in a similar way to the derivation of Eq. \eqref{P_1} in Appendix D: we note that while a collective shift $\Delta_M\propto \mathrm{Re}[G(\omega_p,\mathbf{k}_M,0)]$ exists for modes outside the light cone, their collective width $\Gamma_M\propto \mathrm{Im}[G(\omega_p,\mathbf{k}_M,0)]$ vanishes (here $G(\omega_p,\mathbf{k}_M,0)$ is the lattice Fourier transform of the dipole-dipole kernel) \cite{Efi5}. Nevertheless, the existence of disorder can lead to a non-collective decay component $\gamma_s$ of the stable spin, which may degrade the memory fidelity \cite{GorshkovCavity}.

To conclude, the above considerations establish the mapping to the generic 1D model, Eq. \eqref{model2}, with essentially the same effective parameters $\Gamma$ and $\gamma_{\mathrm{loss}}$ found above for a subwavelength 2D array (Table  \ref{table:1}), and whose scaling from Table  \ref{table:2} were verified numerically by a simple reflectivity calculation.

\section{Discussion: universal approach}
This work introduces an approach for analyzing quantum light-matter interfaces by putting forward the universal role of the reflectivity. The approach roughly consists of two complementary parts: (1) Introducing a minimal 1D scattering model of a quantum interface which is fully characterized by a reflectivity $r_0$: Within the model we found that the efficiencies of quantum applications such as memory and photonic entanglement were equal to $r_0$; (2) Mapping 2D and 3D atom-array interfaces onto the 1D model, finding $r_0$, and hence their efficiencies for the quantum applications mentioned above.

Going forward, this approach opens the prospect for a unified treatment of more applications and platforms. First, consider quantum protocols and tasks beyond those discussed here; say, the generation of photonic 1D cluster states via an array \cite{Bekenstein}. Then, instead of calculating the protocol on the atom-array system, the idea promoted here is to implement it on the simple 1D model (or its suitable variant). The result, in terms of $r_0$ or $C$, can then be related to the required 2D or 3D atom-array system via the mapping in Table I.

Second, consider that one is interested in atom-array interfaces beyond those covered in Table I. A simple example is the inclusion of additional imperfections such as missing atoms or inhomogeneity between atom-like emitters. In this case, one can just calculate the reflectivity classically as demonstrated here, also including these imperfections, thus extracting $r_0$ and the efficiency of the quantum tasks. More generally, going beyond arrays, finding $r_0$ requires first to identify if the mapping to the 1D model is possible or how it can be generalized. This important direction should be pursued based on the considerations presented in Sec. IV.

\begin{acknowledgments}
We acknowledge fruitful discussions with Nir Davidson, Ofer Firstenberg, Darrick Chang and Inbar Shani, and financial support from the Israel Science Foundation (ISF) grant No. 2258/20, the ISF and the Directorate for Defense Research and Development (DDR\&D) grant No. 3491/21, the Center for New Scientists at the Weizmann Institute of Science, the Council for Higher Education (Israel), and QUANTERA (PACE-IN). This research is made possible in part by the historic generosity of the Harold Perlman Family.

\end{acknowledgments}
\appendix
\section{Reflectivity as an efficiency}
Here we provide details on the classical treatment of the 1D scattering problem of Sec. III, highlighting the role of the resonant reflectivity $r_0$ as the efficiency of radiation and absorbtion.
Considering first the radiation from an initially excited dipole in the absence of incident fields,  we solve classically Eq. \eqref{model} for the atomic dipole $P$, with the initial conditions $P(0)=1$ and ${\mathcal{E}}_0=0$ (ignoring vacuum noise in the classical regime), finding
\begin{equation}
P(t)=e^{[i(\delta_p-\Delta)-\frac{\Gamma+\gamma_{\mathrm{loss}}}{2}]t}.
\end{equation}
Inserting this in the equation of the output target field we obtain
\begin{equation}
\mathcal{E}=i\sqrt{\Gamma}e^{[i(\delta_p-\Delta)-\frac{\Gamma+\gamma_{\mathrm{loss}}}{2}]t}.
\end{equation}
The fraction of energy that is emitted to the target mode is then
\begin{eqnarray}
\int_0^\infty|\mathcal{E}|^2dt&=&\frac{\Gamma}{\Gamma+\gamma_{\mathrm{loss}}}=\frac{C}{1+C}=r_0.
\end{eqnarray}
That is, we proved that the radiation efficiency from the dipole to the desired target mode is given by the on-resonance reflectivity $|r(\delta_p=\Delta)|=r_0$.

Next, we turn to the absorption problem, considering a continuous wave (CW) illumination in the target mode and calculating the fraction of power absorbed by the dipole $P$ in steady state. We notice that the interaction term in the dynamical equation \eqref{model} of $P$, can be derived from an effective interaction Hamiltonian $H=-\hbar\sqrt{\Gamma}\mathcal{E}_0 P^\dagger+\mathrm{h.c.}$. This has the form of a force $f=\hbar\sqrt{\Gamma}\mathcal{E}_0$ at a frequency  $\omega_p$ acting on a dipole coordinate $P$. From linear response theory \cite{Landau}, the power dissipated on the dipole is given by  $W=\frac{1}{2}\omega_p\mathrm{Im}[\chi(\omega_p)]|f|^2$, where $\chi$ is the susceptibility of the system defined by $P=\chi f=\chi\hbar\sqrt{\Gamma}\mathcal{E}_0$. To find $\chi$ we solve classically the equation for $P$ in steady state,
\begin{equation}
P=\frac{1}{\frac{\Gamma+\gamma_{\mathrm{loss}}}{2}-i(\delta_p-\Delta)}i\sqrt{\Gamma}\mathcal{E}_0,
\end{equation}
identifying $\chi$ as
\begin{equation}
\chi=\frac{i/\hbar}{\frac{\Gamma+\gamma_{\mathrm{loss}}}{2}-i(\delta_p-\Delta)}.
\end{equation}
Therefore, on resonance $\delta_p=\Delta$, the energy absorbed by the system is
\begin{equation}
W=\frac{\omega_p}{\hbar}\frac{1}{\Gamma+\gamma_{\mathrm{loss}}}|\hbar\sqrt{\Gamma}\mathcal{E}_0|^2=\frac{\Gamma}{\Gamma+\gamma_{\mathrm{loss}}}\hbar\omega_p|\mathcal{E}_0|^2.
\end{equation}
Now, since the input power is $\hbar\omega_p|\mathcal{E}_0|^2$, we find that the absorbtion efficiency, which is the ratio between the absorbed power and the input power, is
\begin{equation}
\frac{\Gamma}{\Gamma+\gamma_{\mathrm{loss}}}=\frac{C}{1+C}=r_0,
\end{equation}
again given by the resonant reflectivity.

Finally, in order to show that $r_0$ indeed describes the on-resonance reflectivity, we reside to the two-sided model from Eq. \eqref{inout} and solve it classically for a CW input field, $\mathcal{E}_{0,\pm}(z)=\mathcal{E}_{0,\pm}(0)e^{\pm ik_pz}$. Solving for $P$ in steady state, and inserting the solution in Eq. \eqref{inout} for the field, we find, e.g. for the right-propagating component (for $z>0$)
\begin{equation}
\mathcal{E}_{+}(z)e^{-ik_p z}=(1+r)\mathcal{E}_{0,+}(0)+r \mathcal{E}_{0,-}(0),
\end{equation}
with $r=r(\delta_p)$ from Eq. \eqref{r}. That is, the left-going field is reflected with amplitude $r$ and the right-going field is transmitted with amplitude $1+r$, as in a 1D problem with reflectivity $r$ which becomes $r_0$ at resonance $\delta_p=\Delta$.
\section{Quantum memory protocol and efficiency}
In this section we will discuss the quantum memory protocol for optimizing the memory efficiency in a general 1D model. We follow Ref. \cite{GorshkovCavity}, where this problem was solved for an effectively equivalent problem of atoms in a cavity, and show how by controlling the temporal pulse shape of the coupling field the storage and retrieval efficiencies can be optimized to $r_0=\frac{C}{1+C}$.

We assume all atoms initially populate the ground state. We define the storage efficiency $e_s$ (retrieval $e_r$) of a photon pulse of length $T$ ($T_r$), as the ratio between the number of stored excitations (retrieved photons) and the number of incoming photons (stored excitations)
\begin{equation}
e_s=\frac{\langle \hat{S}^\dagger(T) \hat{S}(T)\rangle}{\int_0^T\langle\mathcal{\hat{E}}_{0}^\dagger\mathcal{\hat{E}}_{0}\rangle dt},\quad e_r=\frac{\int_{t_0}^{t_0+T_r}\langle\mathcal{\hat{E}}^\dagger\mathcal{\hat{E}}\rangle dt}{\langle\hat{S}^\dagger(t_0)\hat{S}(t_0)\rangle},
\end{equation}
so that the total efficiency of the whole process, storage + retrieval, is $e_{\text{total}}=e_se_r$.

Beginning with the storage problem, we first define the input pulse shape $h_0(t)$ which is non-zero at $[0,T]$, and normalized according to $\int_0^T|h_0(t)|^2dt=1$. To define $h_0(t)$, we introduce a complete, orthonormal set of functions $\{h_p(t)\}$ which satisfy $\int_0^\infty dth_p^*(t)h_p'(t)=\delta_{p,p'}$ and $\sum_ph_p^*(t)h_p(t')=\delta(t-t')$, and corresponding photon-mode lowering operators $\hat{a}_p=\int_0^\infty dt\mathcal{\hat{E}}_0(t)h_p^*(t)$, such that the quantum field can be written as $\mathcal{\hat{E}}_{0}(t)=\sum_{p}h_p(t)\hat{a}_{p}$. By adiabatically eliminating $\hat{P}$ from Eqs. (\ref{model2}), we obtain the dynamical equation for $\hat{S}(t)$,
\begin{equation}
\frac{d\hat{S}}{dt}=-\bigg[\frac{\Gamma_S}{2}+i(\Delta_S-\delta_{2})\bigg]\hat{S}-\hat{\Omega}_S,
\end{equation}
with
\begin{equation}
\frac{\Gamma_S}{2}+i\Delta_S=\frac{|\Omega|^{2}}{\frac{\Gamma+\gamma_{\mathrm{loss}}}{2}+i\left(\Delta-\delta_{p}\right)},
\end{equation}
\begin{equation}
\hat{\Omega}_S=-\frac{i\Omega^{*}}{\frac{\Gamma+\gamma_{\mathrm{loss}}}{2}+i(\Delta-\delta_{p})}\left(i\sqrt{\Gamma}\mathcal{\hat{E}}_{0}+\hat{F}\right),
\end{equation}
where $\Gamma_S$, $\Delta_S$  and $\hat{\Omega}_S$ are the collective two-photon width, shift and field, respectively. Assuming incident light only in the pulse mode $h_0(t)$, we solve the equations for $\hat{S}$ obtaining
\begin{equation}
e_s=\frac{C}{1+C}\bigg|\int_0^Th_0(t)f(t)dt\bigg|^2,
\end{equation}
with
\begin{equation}
\begin{split}
&f(t)=-\frac{\frac{\Gamma_S}{2}+i\Delta_S}{\Omega(t)}\sqrt{\gamma_{\mathrm{loss}}(1+C)}e^{-\int_t^Td\tau\bigg(\frac{\Gamma_S}{2}+i(\Delta_S-\delta_2)\bigg)}.
\end{split}
\end{equation}
We note that $\int_0^T|f(t)|^2dt\leq1$, with the equality achieved under the condition $\int_0^T\Gamma_Sdt\gg1$. Therefore, when this condition is fulfilled, $\bigg|\int_0^Th_0(t)f(t)dt\bigg|^2$ can be seen as a scalar product between normalized eigenfunctions $h_0(t)$ and $f(t)$, so that $e_s\leq\frac{C}{1+C}$, with the equality achieved when $f(t)=h_0^*(t)$. Extracting $\Omega(t)$ from $f(t)=h_0^*(t)$, we thus get the following dependence of the coupling field pulse shape on the quantum field pulse shape
\begin{equation}
\begin{split}
&\Omega(t)=\frac{-\frac{\gamma_{\mathrm{loss}}+\Gamma}{2}-i(\Delta-\delta_p)}{\sqrt{\gamma_{\mathrm{loss}}+\Gamma}}\times\\&\frac{h_0(t)}{\bigg(\sqrt{\int_0^t|h_0(\tau)|^2d\tau}\bigg)^{\bigg(1+i2\frac{\Delta-\delta_p}{\gamma_{\mathrm{loss}}+\Gamma}\bigg)}}e^{i\delta_2(T-t)}.
\end{split}
\label{pulseshape}
\end{equation}
This is the control pulse shape that maximizes the storage efficiency of the pulse $h_0(t)$, achieving $e_s=\frac{C}{1+C}=r_0$.

To calculate the retrieval efficiency, we similarly define the output pulse shape which we want to retrieve $h_r(t)$ being non-zero at $[t_0,t_0+T_r]$ and normalized according to $\int_{t_0}^{t_0+T_r}|h_r(t)|^2dt=1$. Then, by solving again the dynamical equation, but now with the initial condition $\langle\hat{S}(t_0)\rangle\neq0$, and with an input field which contains only the vacuum, we find that the optimal retrieval is
\begin{equation}
e_r=\frac{C}{1+C}\bigg(1-e^{-\int_{t_0}^{t_0+T_r}\Gamma_Sd\tau}\bigg).
\end{equation}
Again, if $\int_{t_0}^{t_0+T_r}\Gamma_S(\tau)d\tau\gg1$, the retrieval efficiency will be maximized up to $e_r=\frac{C}{1+C}=r_0$. As with the calculation performed for the storage process, we can determine the control pulse shape that will maximize the retrieval efficiency. We obtain the (conjugate) time reverse expression as in \eqref{pulseshape}. That is, the optimal control that retrieves the photons into a mode $h_r(t)$ is just the (conjugate) time reverse of the control needed to optimally store an input pulse with a reversed time shape. The reason is that the storage and retrieval are reversed symmetric processes of each other \cite{GorshkovCavity}.
\section{Mapping the collective system to the 1D model}
In this section we discuss in detail the mapping of the field equation of the collective system to the 1D model. Performing the projection \eqref{transformed_E} on the field equation \eqref{E_r}, we have
\begin{equation}
\begin{split}
&\hat{E}_u(z)=\hat{E}_{u,0}(z)+\\&
\frac{1}{\sqrt{A_u}}\frac{\omega_{p}^{2}d}{\epsilon_{0}c^{2}}\sum_{n}\hat{\sigma}_{ge,n}\int_{-\infty}^{\infty}G(\omega_{p},\mathbf{r}_{\perp}-\mathbf{r}^{\perp}_{n},z-z_n)u^*(\mathbf{r}_{\perp})d\mathbf{r}_{\perp}.
\end{split}
\label{in_out}
\end{equation}
Writing the Green's function in the in-plane momentum expansion \cite{Efi5}
\begin{equation}
\begin{split}
&G(\omega_{p},\mathbf{r}_{\perp}-\mathbf{r}^{\perp}_{n},z-z_n)=\\&=\frac{i}{8\pi^{2}}\int_{-\infty}^{\infty}d\textbf{k}_\perp\left(1-\frac{|\textbf{k}_\perp\cdot\textbf{e}_{d}|^{2}}{k_p^{2}}\right)e^{i\textbf{k}_\perp\cdot\left(\mathbf{r}_{\perp}-\mathbf{r}^{\perp}_{n}\right)}\frac{e^{ik_{z}|z-z_n|}}{k_{z}},
\end{split}
\end{equation}
with $k_p=2\pi/\lambda$ and $k_z=\sqrt{k_p^2-|\textbf{k}_\perp|^2}$, and inserting it into Eq. \eqref{in_out}, we obtain
\begin{equation}
\begin{split}
&\hat{E}_u(z)=\hat{E}_{u,0}(z)+\frac{1}{\sqrt{A_u}}\frac{i}{4\pi}\frac{\omega_{p}^{2}d}{\epsilon_{0}c^{2}}\times\\&\sum_{n}\hat{\sigma}_{ge,n}\int_{-\infty}^{\infty}d\textbf{k}_\perp\left(1-\frac{|\textbf{k}_\perp\cdot\textbf{e}_{d}|^{2}}{k_p^{2}}\right)\frac{e^{ik_{z}|z-z_n|}}{k_{z}}e^{-i\textbf{k}_\perp\cdot\mathbf{r}^{\perp}_{n}}\tilde{u}^*(\textbf{k}_\perp),
\end{split}
\end{equation}
where $\tilde{u}(\textbf{k}_\perp)=\frac{1}{2\pi}\int_{-\infty}^{\infty}e^{-i\textbf{k}_\perp\cdot\mathbf{r}_{\perp}}u(\mathbf{r}_{\perp})d\mathbf{r}_{\perp}$ is the Fourier transform of the transverse mode. Now, we assume the paraxial approximation in which the width of the spatial transverse mode $u(\mathbf{r}_{\perp})$, is much larger than the wavelength, $w\gg\lambda$, and we consider propagation distances within its Rayleigh range $z<z_R=\pi w^2/\lambda$, so diffraction of the mode can be neglected. In Fourier space this means that $\tilde{u}(\textbf{k}_\perp)$ is narrow so that it can be approximated as existing only at the central frequency, yielding $k_{z}=\sqrt{k_p^{2}-|\textbf{k}_\perp|^{2}}\approx k_p$, and $\frac{|\textbf{k}_\perp\cdot\textbf{e}_{d}|^{2}}{k_p^{2}}\ll1$; the requirement $z<z_R$ further allows to approximate the phase factor $e^{ik_z|z-z_n|}\approx e^{ik_p|z-z_n|}$.  Under these approximations, we obtain [by transforming $\tilde{u}^*(\textbf{k}_\perp)$ back to real space],
\begin{equation}
\begin{split}
&\hat{E}_u(z)=\hat{E}_{u,0}(z)+\frac{idk_p}{2\epsilon_{0}\sqrt{A_u}}\sum_{n}u^*(\mathbf{r}^{\perp}_n)e^{ik_p|z-z_n|}\hat{\sigma}_{ge,n}.
\label{map_E}
\end{split}
\end{equation}
Decomposing $\hat{E}_u(z)$ into $\hat{E}_{u}^+(z)$ and $\hat{E}_{u}^-(z)$ which are the right- and left-propagating fields [including only $k_z>0$ or $k_z<0$, respectively], sampled at $z>z_{\mathrm{max}}$ and $z<z_{\mathrm{min}}$ respectively ($z_n\in[z_{\mathrm{min}},z_{\mathrm{max}}]$), we arrive at Eq. \eqref{in_out_transformed1} of the main text.
\section{Mapping the 2D array to the 1D model}
In the following, we discuss the mapping of the 2D array to the 1D model. For the mapping of the field equation, we consider Eq. \eqref{map_E} for the 2D array ($z_n=0,\forall n$) by using the definition (\ref{transform_P}) for $\hat{P}$, obtaining
\begin{equation}
\hat{E}_u(z)=\hat{E}_{u,0}(z)+i\frac{\hbar}{d}\frac{\sqrt{\eta}\Gamma_0}{2}e^{ik_p|z|}\frac{a}{\sqrt{A_u}}\hat{P}_u.
\end{equation}
By decomposing the field into its right and left propagating components, we arrive at equation \eqref{in_out_transformed}, which we then cast to the form of the input-output equation of the 1D model, as discussed in the main text.

For the mapping of the atomic equations we transform equations \eqref{sigma_ge_real}-\eqref{sigma_gs_real} (with $z_n=0,\forall n$), according to \eqref{transform_P}-\eqref{transform_S}. The equation for $\hat{S}_u$ transforms trivially, and for $\hat{P}_u$ we obtain
\begin{equation}
\begin{split}
&\frac{d\hat{P}_u}{dt}=-\left(\frac{\gamma_{s}}{2}-i\delta_{p}\right)\hat{P}_u+i\Omega\hat{S}_u+\\&+\frac{id}{\hbar}\frac{a}{\sqrt{\eta}}\sum_{n}\hat{E}_{0}(\mathbf{r}^{\perp}_{n},0)u^*(\mathbf{r}^{\perp}_n)+\hat{F}_u\\&+\frac{i}{\hbar}\frac{d^2\omega_{p}^{2}}{\epsilon_{0}c^{2}}\frac{a}{\sqrt{\eta}}\sum_{n}\sum_{m}G(\omega_{p},\mathbf{r}^{\perp}_n-\mathbf{r}^{\perp}_{m},0)u^*(\mathbf{r}^{\perp}_n)\hat{\sigma}_{ge,m},
\end{split}
\label{transformation}
\end{equation}
with $\hat{F}_u=\frac{a}{\sqrt{\eta}}\sum_{n}\hat{F}_nu^*(\mathbf{r}^{\perp}_n)$. We introduce a complete basis $\{u_{\beta}(\mathbf{r}_{\perp})\}$ which spans the functions space of the $xy$ plane, which is where our transverse mode $u(\mathbf{r}_{\perp})$ lives, such that $\sum_{\beta}u_{\beta}(\mathbf{r}_{\perp})u_{\beta}^*(\mathbf{r}_{\perp}')=\delta(\mathbf{r}_{\perp}-\mathbf{r}_{\perp}')$. In case of a Gaussian mode, the basis is the Hermite Gauss modes and $u(\mathbf{r}_{\perp})\equiv u_0(\mathbf{r}_{\perp})$ is the Gaussian mode  $\beta=0$. With this basis the photon field can be written as an inverse transformation $\hat{E}(\mathbf{r}_{\perp},z)=\sum_{\beta}\sqrt{A_{\beta}}\hat{E}_{\beta}(z)u_{\beta}(\mathbf{r}_{\perp})$,
so we have
\begin{equation}
\begin{split}
&\frac{id}{\hbar}\frac{a}{\sqrt{\eta}}\sum_{n}\hat{E}_{0}(\mathbf{r}^{\perp}_{n},0)u^*(\mathbf{r}^{\perp}_n)=\\&=\frac{id}{\hbar}\frac{a}{\sqrt{\eta}}\sqrt{A_u}\hat{E}_{u,0}(0)\sum_{n}u^*(\mathbf{r}^{\perp}_n)u(\mathbf{r}^{\perp}_n)+\\&+\frac{id}{\hbar}\frac{a}{\sqrt{\eta}}\sum_{\beta\neq0}\sqrt{A_{\beta}}\hat{E}_{\beta,0}(0)\sum_{n}u_{\beta}(\mathbf{r}^{\perp}_n)u^*(\mathbf{r}^{\perp}_n)=\\
&=i\sqrt{\eta\Gamma_0}\mathcal{\hat{E}}_{u,0}(0)+\hat{F}_\eta.
\end{split}
\end{equation}
Here, we used the fact that for the paraxial mode $w \gg a$ we can approximate $\sum_{n}\approx\frac{1}{a^2}\int_{L_a^2}d\mathbf{r}^{\perp}_n$. We assume that only the (Gaussian) $u(\mathbf{r}_{\perp})$ mode, with $\beta=0$, is populated, and the quantum noise operator $\hat{F}_\eta=\frac{id}{\hbar}\frac{a}{\sqrt{\eta}}\sum_{\beta\neq0}\sqrt{A_{\beta}}\hat{E}_{\beta,0}(0)\sum_{n}u_{\beta}^*(\mathbf{r}^{\perp}_n)u(\mathbf{r}^{\perp}_n)$ accounts for the vacuum fluctuations of other modes, which corresponds to unwanted emissions due to the mismatch in the overlap between the array and the beam wavefront.

For the last line in \eqref{transformation} we use the inverse transformation of the transverse mode in Fourier space $u(\mathbf{r}_{\perp})=\frac{1}{2\pi}\int \tilde{u}(\textbf{k}_\perp)e^{i\textbf{k}_\perp\cdot\mathbf{r}_{\perp}}d\textbf{k}_\perp$ obtaining
\begin{equation}
\begin{split}
&\frac{a}{\sqrt{\eta}}\sum_{n}\sum_{m}G(\omega_{p},\mathbf{r}^{\perp}_n-\mathbf{r}^{\perp}_{m},0)\hat{\sigma}_{ge,m}u^*(\mathbf{r}^{\perp}_n)=\\
&=\frac{a}{\sqrt{\eta}}\sum_{m}\hat{\sigma}_{ge,m}\frac{1}{2\pi}\int d\textbf{k}_\perp \tilde{u}^*(\textbf{k}_\perp)G_m(\omega_p,\textbf{k}_\perp,0)e^{-i\textbf{k}_\perp\cdot\mathbf{r}^{\perp}_m},
\end{split}
\end{equation}
where we defined $G_m(\omega_p,\textbf{k}_\perp,0)=\sum_{n}G(\omega_{p},\mathbf{r}^{\perp}_n-\mathbf{r}^{\perp}_{m},0)e^{-i\textbf{k}_\perp\cdot(\mathbf{r}^{\perp}_n-\mathbf{r}^{\perp}_m)}$. We assume that the array is larger than the wavelength $L_a\gg\lambda\rightarrow\sqrt{N}\gg\lambda/a$, so each atom in the ``bulk'' (not at the edges) effectively feels interactions of infinite array (noting the Green's function oscillates with spatial frequency $1/\lambda$). In addition, this assumption yields the condition $\sqrt{N}\gg1$, so most of the atoms are in the bulk, and edge atoms are negligible in describing collective dipoles. Under these conditions we can approximate $G_m(\omega_p,\textbf{k}_\perp,0)\approx G(\omega_p,\textbf{k}_\perp,0)=\sum_{n\in \mathrm{infinite}}G(\omega_{p},\mathbf{r}^{\perp}_n,0)e^{-i\textbf{k}_\perp\cdot\mathbf{r}^{\perp}_n}$, where $G(\omega_p,\textbf{k}_\perp,0)$ is that of an infinite array. Then, assuming that the width of $\tilde{u}^*(\textbf{k}_\perp)$ is narrower than that of the Green's function $G(\omega_p,\textbf{k}_\perp,0)$, the latter is effectively sampled at $\textbf{k}_\perp=0$, obtaining
\begin{equation}
\begin{split}
&=\frac{a}{\sqrt{\eta}}G(\omega_p,0,0)\sum_{m}\hat{\sigma}_{ge,m}u^*(\mathbf{r}^{\perp}_m)=G(\omega_p,0,0)\hat{P}_u.
\end{split}
\end{equation}
The Green's function $G(\omega_p,0,0)$ of a subwavelength ordered array is proportional the collective emission rate and detuning of the zero-momentum mode, which is the coupling to a normal incidence light \cite{Efi5}
\begin{equation}
\frac{i}{\hbar}\frac{d^2\omega_{p}^{2}}{\epsilon_{0}c^{2}}G(\omega_p,0,0)=-\frac{\Gamma_0}{2}-i\Delta_0
\label{Green_Gamma}.
\end{equation}
Putting everything together, equation \eqref{transformation} becomes
\begin{eqnarray}
\frac{d\hat{P}_u}{dt}&=&-\left[\frac{\Gamma_0+\gamma_{s}}{2}+i(\Delta_0-\delta_{p})\right]\hat{P}_u+i\Omega\hat{S}_u\nonumber\\&+&i\sqrt{\eta\Gamma_0}\mathcal{\hat{E}}_{u,0}(0)+\hat{F}_\eta+\hat{F}_u,
\end{eqnarray}
which is Eq. \eqref{P_1} in the main text, from which it is shown how the mapping to the 1D model is completed.
\section{Mapping the 3D array to the 1D model}
Here, we discuss the mapping of the 3D multilayer array to the 1D model. The collective dipole \eqref{transform_P_3D} now has an additional index $n_z$ indicating its layer $n_z$, and its dynamics are given by
\begin{equation}
\begin{split}
&\frac{d\hat{P}_{u,n_z}}{dt}=-\left(\frac{\gamma_{s}}{2}-i\delta_{p}\right)\hat{P}_{u,n_z}+i\Omega\hat{S}_{u,n_z}+\\&+i\sqrt{\eta\Gamma_0}\mathcal{\hat{E}}_{u,0,n_z}+\hat{F}+\frac{i}{\hbar}\frac{d^2\omega_{p}^{2}}{\epsilon_{0}c^{2}}\frac{a}{\sqrt{\eta}}\times\\&\sum_{m_z=0}^{N-1}\sum_{n}\sum_{m}G(\omega_{p},\mathbf{r}^{\perp}_n-\mathbf{r}^{\perp}_{m},a_z(n_z-m_z))u^*(\mathbf{r}^{\perp}_n)\hat{\sigma}_{ge,m,m_z},
\end{split}
\end{equation}
where the input field is $i\sqrt{\eta\Gamma_0}\mathcal{\hat{E}}_{u,0,n_z}=\frac{id}{\hbar}\frac{\sqrt{A_u}}{a}\sqrt{\eta}\hat{E}_{u,0}(a_zn_z)$. For the last line we follow the same procedure as we did for the single layer case, obtaining
\begin{equation}
\begin{split}
&\frac{a}{\sqrt{\eta}}\sum_{m_z=0}^{N-1}\sum_{n}\sum_{m}\\&G(\omega_{p},\mathbf{r}^{\perp}_n-\mathbf{r}^{\perp}_{m},a_z(n_z-m_z))u^*(\mathbf{r}^{\perp}_n)\hat{\sigma}_{ge,m,m_z}\\
&=\sum_{m_z=0}^{N-1}\frac{a}{\sqrt{\eta}}\sum_{m\leq N}\hat{\sigma}_{ge,m,m_z}\frac{1}{2\pi}\int d\textbf{k}_\perp \tilde{u}^*(\textbf{k}_\perp)e^{-i\textbf{k}_\perp\cdot\mathbf{r}^{\perp}_m}\\&G_m(\omega_p,\textbf{k}_\perp,a_z(n_z-m_z)),
\end{split}
\end{equation}
where we defined for any two $n_z,m_z$ layers $G_m(\omega_p,\textbf{k}_\perp,a_z(n_z-m_z))=\sum_{n\le N}G(\omega_{p},\mathbf{r}^{\perp}_n-\mathbf{r}^{\perp}_{m},a_z(n_z-m_z))e^{-i\textbf{k}_\perp\cdot(\mathbf{r}^{\perp}_n-\mathbf{r}^{\perp}_m)}$. Now, as before we assume $L_a\gg\lambda$ and $\sqrt{N}\gg1$, so that the Green's function between any two finite layers can be approximated by the Green's function between two  infinite layers, $G_m(\omega_p,\textbf{k}_\perp,a_z(n_z-m_z))\approx G(\omega_p,\textbf{k}_\perp,a_z(n_z-m_z))$. Then, we assume the paraxial approximation in which the width of the spatial transverse mode $\tilde{u}^*(\textbf{k}_\perp)$ is narrow and $a_z|n_z-m_z|<z_R,L_a^2/\lambda$, so that the Green's function is sampled at $\textbf{k}_\perp=0$, obtaining
\begin{equation}
\begin{split}
&=\sum_{m_z=0}^{N-1}G(\omega_p,0,a_z(n_z-m_z))\frac{a}{\sqrt{\eta}}\sum_{m}u^*(\mathbf{r}^{\perp}_m)\hat{\sigma}_{ge,m,m_z}=\\&=\sum_{m_z=0}^{N-1}G(\omega_p,0,a_z(n_z-m_z))\hat{P}_{u,m_z}.
\end{split}
\end{equation}
The scattered Green's function at position $z$ can be written as \cite{Efi5}
\begin{equation}
\begin{split}
&\frac{i}{\hbar}\frac{d^2\omega_{p}^{2}}{\epsilon_{0}c^{2}}G(\omega_p,0,z)=\\&=-\frac{\Gamma_0}{2}\sum_{(m_x,m_y)}\frac{1-\frac{\lambda^2}{a^2}|(m_x,m_y)\cdot\textbf{e}_d|^2}{\sqrt{1-\frac{\lambda^2}{a^2}(m_x^2+m_y^2)}}e^{ik_p\sqrt{1-\frac{\lambda^2}{a^2}(m_x^2+m_y^2)}|z|},
\end{split}
\end{equation}
so we have
\begin{equation}
\begin{split}
&\frac{i}{\hbar}\frac{d^2\omega_{p}^{2}}{\epsilon_{0}c^{2}}\sum_{m_z=0}^{N-1}G(\omega_p,0,a_z(n_z-m_z))\hat{P}_{u,m_z}=\\
&-\frac{\Gamma_0}{2}\sum_{m_z=0}^{N-1}\sum_{(m_x,m_y)}\frac{1-\frac{\lambda^2}{a^2}|(m_x,m_y)\cdot\textbf{e}_d|^2}{\sqrt{1-\frac{\lambda^2}{a^2}(m_x^2+m_y^2)}}\times\\&e^{ik_p\sqrt{1-\frac{\lambda^2}{a^2}(m_x^2+m_y^2)}a_z|n_z-m_z|}\hat{P}_{u,m_z}.
\end{split}
\end{equation}
We separate from the sum the self-interaction term $m_z=n_z$, obtaining
\begin{equation}
\begin{split}
&=-\bigg(\frac{\Gamma_0}{2}+i\Delta_0\bigg)\hat{P}_{u,n_z}-\sum_{m_z\neq n_z}^{N-1}D_{n_z,m_z}\hat{P}_{u,m_z},
\end{split}
\end{equation}
with $D_{n_z,m_z}$ from Eq. \eqref{D} in the main text.
Putting this back into the dynamical equation we arrive at Eq. \eqref{P_nz} of the main text.

For the field equation we start with Eq. \eqref{map_E}, with an additional summation over the layers, obtaining
\begin{equation}
\begin{split}
&\hat{E}_u(z)=\hat{E}_{u,0}(z)+\frac{idk_p}{2\epsilon_{0}\sqrt{A_u}}\sum_{n_z=0}^{N_z-1}e^{ik_p|z-a_zn_z|}\frac{\sqrt{\eta}}{a}\hat{P}_{u,n_z}.
\label{map_E3d}
\end{split}
\end{equation}
The difference between the single 2D lattice case is that now each layer has an extra relative phase proportional to its position along the $z$ axis. If $a_z$ is a half integer number of the wavelength, this phase is identical for all layers up to a sign, obtaining the same sign for $2a_z/\lambda$ even, or alternating signs for $2a_z/\lambda$ odd. In either case the phase of each layer matches the phase of the collective dipole defined by Eq. \eqref{P_u}. Now, we define the symmetric field $\mathcal{\hat{E}}_u(z)=\frac{1}{\sqrt{2}}\bigg[\mathcal{\hat{E}}_u^+(z)+\mathcal{\hat{E}}_u^-(-z)\bigg]e^{-ik_pz}$ ($z>0$), where $\mathcal{\hat{E}}_u^{\pm}(z)$ are the right- and left-propagating fields, sampled at $z>a_z(N_z-1)$ and $z<0$ respectively, as defined in Eqs. (\ref{in_out_transformed1}) and \eqref{in_out_transformed}. With these considerations the input-output relation for $\mathcal{\hat{E}}_u(z)$ reduces to
\begin{equation}
\mathcal{\hat{E}}_u(z)=\mathcal{\hat{E}}_{u,0}(z)+i\sqrt{\eta N_z\Gamma_0}\hat{P}_u,
\end{equation}
which is Eq. \eqref{E_3D} in the main text.
\section{Correction $\Delta'$ to the 3D array collective shift}
\begin{figure}[t!]
  \begin{center}
    \includegraphics[width=\columnwidth]{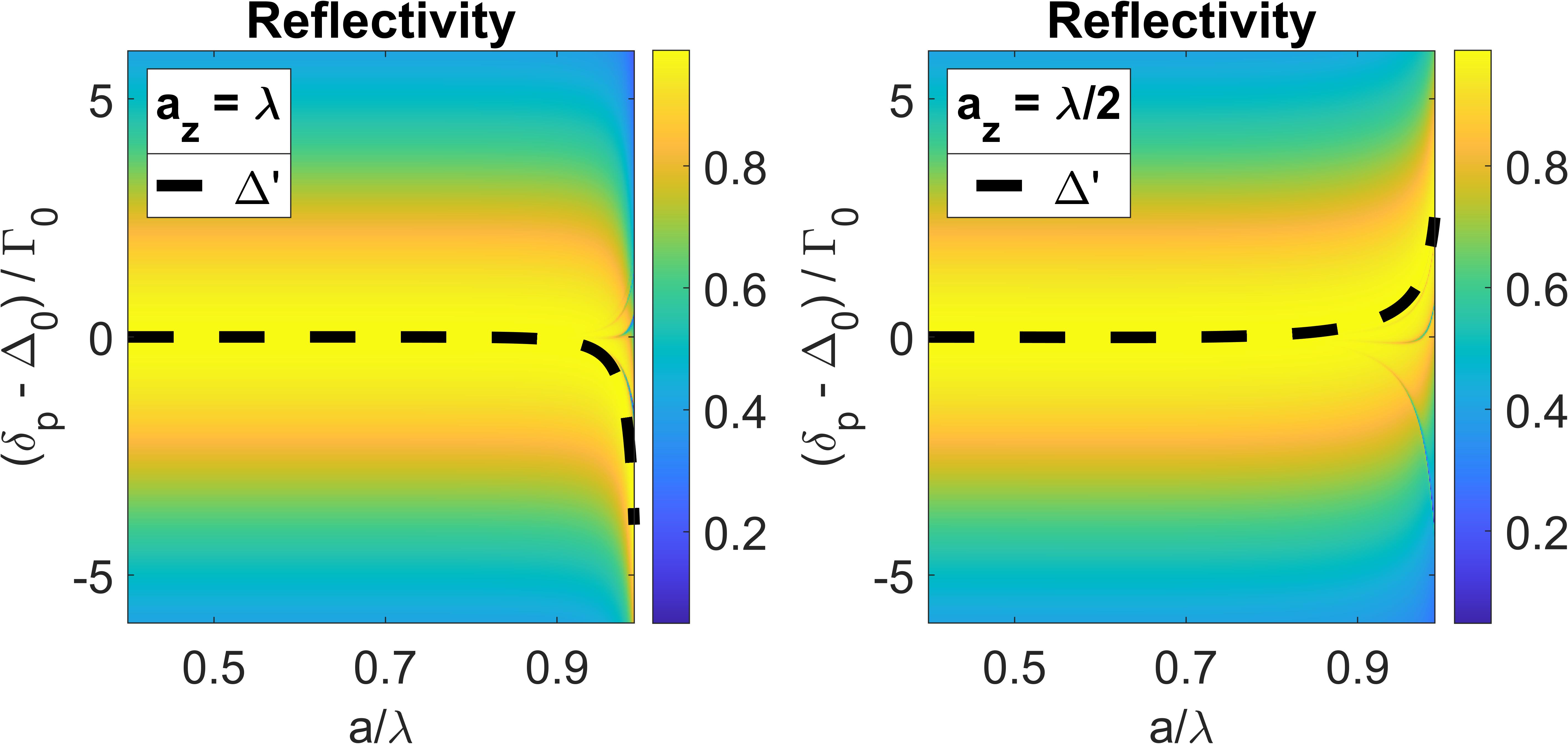}
    \caption{\small{Intensity reflection coefficient $R$ of a 3D multilayered array, as a function of detuning $\delta_{p}-\Delta_{0}$ and the ratio $a/\lambda$, for (a) $a_{z}=\lambda$, and (b) $a_{z}=\lambda/2$. The detuning correction $\Delta'$ (dashed line) fits the maximal reflectivity. The reflectivity was calculated for an array with $N_{z}=10$, $\eta=1$ and $\gamma_{\text{loss}}=0.05\Gamma_{0}$.
    }} \label{first_order_correction}
  \end{center}
\end{figure}
Here we elaborate on the estimation of the correction $\Delta'$ to the collective shift, appearing in Eq. \eqref{P_3D_final} for the collective dipole $\hat{P}_u$ of a 3D array. As explained in the main text, this shift comes from a perturbative treatment of the evanescent-field component, $\varepsilon_{n_{z}m_{z}}$, of the interaction between layers. To first order in perturbation theory, the correction to the relevant collective eigenmode from Eq. \eqref{P_u}, $\left|v\right\rangle =\frac{1}{\sqrt{N_z}}\left(1,e^{-ik_{p}a_{z}},\dots,e^{-ik_{p}a_{z}\left(N_{z}-1\right)}\right)$ is given as usual by the matrix element $\left\langle v|\varepsilon|v\right\rangle $,
\begin{equation}
\Delta'=\frac{1}{N_{z}}\sum_{n_{z}\neq m_z}e^{ik_{p}a_{z}\left(n_{z}-m_{z}\right)}\varepsilon_{n_{z},m_{z}}.
\label{Delta'}
\end{equation}
We now verify that this approximation is sufficiently accurate for relevant cases. To this end, we consider the calculation of the 3D array reflectivity in two ways. From the mapping to the generic 1D model as in Eq. \eqref{P_3D_final}, we readily predict that the optimal reflectivity is obtained at a resonance \emph{shifted} by $\Delta'$, i.e. for $\delta_p=\Delta_0+\Delta'$, using $\Delta'$ from Eq.\eqref{Delta'}. This is compared to an exact calculation which does not rely on the first-order perturbative approximation of Eq. \eqref{Delta'} and the subsequent mapping to the generic 1D model: Beginning with Eq. \eqref{P_nz} with the full inter-layer kernel from \eqref{mixing_term} (including $\varepsilon_{n_{z}m_{z}}$), we classically solve for the dipole $P_{u,n_z}$ of the layer ${n_z}$ given a CW input field, by a simple matrix inversion performed numerically. Then, plugging this solution into Eq. \eqref{map_E} we find the total field and the reflectivity. The results of this exact numerical calculation are presented in Fig. \ref{first_order_correction}
as a function of the detuning and the ratio $a/\lambda$ and for both types of phase-matched collective dipoles, i.e. $a_{z}=\lambda$ and $a_{z}=\lambda/2$. Maximal reflectivity is indeed observed to overlap the curve of the $\Delta'(a/\lambda)$ calculated from Eq. \eqref{Delta'} as a function of $a/\lambda$, as predicted by the approximate generic 1D model with the shift $\Delta'$. As a relevant example, consider the lattice spacing $a/\lambda=0.68$ corresponding to a typical optical lattice experiment \cite{Rui}: in this case, we see in Fig. \ref{first_order_correction} excellent agreement between the position of the maxima of the exact result and the corresponding value of $\Delta'$. To conclude, we find that the mapping to the generic 1D model as per Eq. \eqref{P_3D_final} is valid for the regions of interest, using the small correction $\Delta'$.

\end{document}